\documentclass{lmcs} 
\pdfoutput=1

\usepackage{lastpage}
\lmcsdoi{21}{4}{26}
\lmcsheading{}{\pageref{LastPage}}{}{}%
{Jun.~19,~2024}{Dec.~05,~2025}{}


\usepackage[utf8]{inputenc}
\usepackage[english]{babel}

\usepackage[footnotesize]{caption}
\usepackage{graphicx}
\usepackage{amsmath}
\usepackage{booktabs}
\usepackage{etex}
\usepackage{amssymb}
\usepackage[T1]{fontenc}
\usepackage{amsfonts}
\usepackage{dsfont}
\usepackage{stmaryrd}
\usepackage{alltt}
\usepackage[all]{xy}
\usepackage{mathrsfs}
\usepackage{array}
\usepackage{tabularx}
\usepackage{wasysym}
\usepackage{hyperref}
\usepackage[all]{xy}
\hypersetup{
  bookmarksnumbered = {true},
  naturalnames      = {true},
  colorlinks        = {true},
  linkcolor         = black,
  urlcolor          = black,
  citecolor         = black
}
\usepackage{cleveref}

\Crefname{algorithm}{Alg.}{Algs.}
\Crefname{defi}{Def.}{Defs.}
\Crefname{figure}{Fig.}{Figs.}
\Crefname{prop}{Prop.}{Props.}
\Crefname{rem}{Remark}{Remarks}
\Crefname{equation}{Eq.}{Eqs.}
\Crefname{thm}{Thm.}{Thms.}
\Crefname{lem}{Lemma}{Lemmas}
\Crefname{exa}{Ex.}{Exs.}
\Crefname{section}{Section}{Sections}
\Crefname{fact}{Fact}{Facts}
\Crefname{appendix}{App.}{App.}
\Crefname{table}{Tab.}{Tabs.}

\usepackage{longtable}
\usepackage{tikz}
\usetikzlibrary{arrows,calc,patterns,positioning,shapes}
\usetikzlibrary{decorations.pathmorphing}

\usepackage{amscd}
\usepackage{xargs}
\usepackage{relsize}
\usepackage{xifthen}
\usepackage{xspace}
\usepackage{mathtools}
\usepackage{textgreek}
\usepackage{proof}
\usepackage[textwidth=0.89in,textsize=scriptsize]{todonotes}
\usepackage{cases}
\usepackage{multicol}

\usepackage{prooftrees}

\usepackage[nomargin,inline,marginclue,draft]{fixme}

\usepackage{amsbsy}

\usepackage{proof}
\usepackage{amsthm}

\usepackage[linesnumbered,ruled,vlined]{algorithm2e}

\usepackage{xcolor}

\usepackage{localdefs}

\begin{document}

\title[Data-Aware Hybrid Tableaux]{Data-Aware Hybrid Tableaux}

\author[C.~Areces]{Carlos Areces\lmcsorcid{0000-0001-7845-8503}}[a,b]
\author[V.~Cassano]{Valentin Cassano\lmcsorcid{0000-0001-5904-3038}}[a,b,c]
\author[R.~Fervari]{Raul Fervari\lmcsorcid{0000-0003-0360-0725}}[a,b]

\address{Universidad Nacional de C\'ordoba, Argentina}	
\email{carlos.areces@unc.edu.ar, rfervari@unc.edu.ar}  

\address{Consejo Nacional de Investigaciones Cient\'ificas y T\'ecnicas (CONICET), Argentina}	

\address{Universidad Nacional de R\'io Cuarto, Argentina}
\email{valentin@dc.exa.unrc.edu.ar}










\maketitle

\begin{abstract}
  Labelled tableaux have been a traditional approach to define satisfiability checking procedures for Modal Logics. In many cases, they can also be used to obtain tight complexity bounds and lead to efficient implementations of reasoning tools. More recently, 
  it has been shown that the expressive power provided by the operators characterizing Hybrid Logics (nominals and satisfiability modalities) can be used to \emph{internalize} labels, leading to well-behaved inference procedures for fairly expressive logics. The resulting procedures are attractive because they do not use external mechanisms outside the language of the logic at hand, and have good logical and computational properties.
  
  Many tableau systems based on Hybrid Logic have been investigated, with more recent efforts concentrating on Modal Logics that support data comparison operators.
  Here, we introduce an internalized tableau calculus for XPath, arguably one of the most prominent approaches for querying semistructured data.  More precisely, we define data-aware tableaux for XPath featuring data comparison operators and enriched with nominals and the satisfiability modalities from Hybrid Logic.  We prove that the calculus is sound, complete and terminating. Moreover, we show that tableaux can be explored in polynomial space, therefore establishing that the satisfiability problem for the logic is \pspace-complete. Finally, we explore different extensions of the calculus, in particular how to handle data trees and other frame classes.
\end{abstract}

\section{Introduction}
\label{sec:intro}

In many data-centric applications --such as those handling large volumes of web or medical information-- explicit reasoning about both structure and content is essential. Classical relational databases often fall short in expressiveness for such use cases, leading to the development of \emph{semi-structured data models}, typically based on labeled trees or graphs~\cite{Bune:semi97,Abit:data99}. A prominent example of such semi-structured data models is XML (eXtensible Markup Language), which encodes structural information via labels from a finite alphabet, and data content using values drawn from an infinite domain. XML documents are commonly queried using XPath, a standard language widely used in tools like XSLT~\cite{xslt:w3c} and XQuery~\cite{xquery:w3c}.  

XPath is, fundamentally, a general purpose language for
addressing, searching, and matching pieces of an XML document. It is an open standard
World Wide Web Consortium (W3C) Recommendation~\cite{xpath:w3c}.
Core-XPath, the fragment of XPath 1.0 that captures its navigational capabilities~\cite{GKP05}, can express structural properties (e.g., tag names, tree traversal), but not conditions involving the actual data values. Logically, Core-XPath corresponds to a classical modal logic~\cite{blackburn2001modal,blackburn06}, and its satisfiability problem is decidable, even under DTD constraints~\cite{M04,BFG08}. Its expressive power has been precisely characterized: it is equivalent to two-variable first-order logic (FO$^2$) on trees~\cite{MdR05} and it is strictly less expressive than PDL with converse~\cite{BK08}. Sound and complete axiomatizations are available in~\cite{CateM09,cateLM10}.

However, Core-XPath’s inability to compare data values across nodes limits its applicability, notably preventing the expression of joins --an essential construct in query languages. To address this, the logic \xpathd (also known as Core-Data-XPath~\cite{BojanczykMSS09}) extends Core-XPath with data-aware features, allowing for expressions such as $\tup{\alpha =_{\textsf{c}} \beta}$ or $\tup{\alpha \neq_{\textsf{c}} \beta}$, where $\alpha$ and $\beta$ are path expressions navigating the XML tree via axes like child, descendant, or sibling. Such expressions enable comparisons of data values for \textsf{c} in nodes reachable via different paths, allowing for significantly richer queries.

By way of example, an expression of the form $\tup{\alpha =_{\textsf{c}} \beta}$ is true at a node $x$ in a data tree if there exist two nodes reachable from $x$ via $\alpha$ and $\beta$ paths respectively, such that the data values for \textsf{c} of $y$ and $z$ are equal. This situation is illustrated in~\Cref{fig:datatree} depicting how the expression {\em ``there is
a one-step descendant via $\textsf{e}$ and a two-steps descendant via $\textsf{e}$ with the same data
value for \textsf{price}''} (written $\tup{\textsf{e} =_{\textsf{price}} \textsf{e}\textsf{e}}$) holds at $x$, given the presence of $u$ and~$z$. The
expression {\em ``there are two two-steps descendant with different data value for \textsf{price}''} (written $\tup{\textsf{e}\textsf{e} \neq_{\textsf{price}} \textsf{e}\textsf{e}}$)
is also true at $x$, because $v$ and $w$ have different data for \textsf{price}.


%

\begin{figure}[h!]	
	
		\begin{tikzpicture}[
			>=latex
			]
			
			\node[draw, shape = rectangle, rounded corners, minimum size = 7mm] 
			   (a) at (0,0) {};
			\node[draw, shape = rectangle, rounded corners, minimum size = 7mm] 
			   (b) [below left=of a] {};
			\node[draw, shape = rectangle, rounded corners, minimum size = 7mm] 
			   (c) [below right=of a] {\textsf{price}=\$10};
			\node[draw, shape = rectangle, rounded corners, minimum size = 7mm] 
			   (d) [below left= of b] {\textsf{price}=\$10};
			\node[draw, shape = rectangle, rounded corners, minimum size = 7mm] 
			   (e) [below=of b] {\textsf{price}=\$10};
			\node[draw, shape = rectangle, rounded corners, minimum size = 7mm] 
			   (f) [below right=of b] {\textsf{price}=\$20};
			
			\draw[->] (a) -- (b) node[midway, left] {\textsf{e}};
			\draw[->] (a) -- (c) node[midway, left] {\textsf{e}};
			\draw[->] (b) -- (d) node[midway, above] {\textsf{e}};
			\draw[->] (b) -- (e) node[midway, right] {\textsf{e}};
			\draw[->] (b) -- (f) node[midway, above] {\textsf{e}};
		
		\begin{scope}[nodes = {right = 11pt}]
			\node[draw=none]  at (a) {$x$};
			\node[draw=none]  at (b) {$y$};
		\end{scope}

		\begin{scope}[nodes = {below = 11pt}]
			\node[draw=none]  at (c) {$z$};
			\node[draw=none]  at (d) {$u$};
			\node[draw=none]  at (e) {$v$};
			\node[draw=none]  at (f) {$w$};
		\end{scope}
			
		\end{tikzpicture}

	\caption{An example of a data tree.}\label{fig:datatree}
\end{figure}

The language of $\xpathd$ allows us to compare data values at the end of a
path, by equality or inequality. However, it does not grant access
to the concrete data value of nodes. This makes it possible to work with an abstraction of data trees:
instead of having concrete data values in each node, we have an
equivalence relation between nodes linking nodes with the same value. In the data tree from~\Cref{fig:datatree}, the relation consists of four equivalence
classes: $\{u,v,z\}$, $\{w\}$, $\{x\}$ and $\{y\}$. This approach treats uniformly nodes with unique data values of a certain type and nodes where data of that type is undefined (see~\cite{arec:data23} for a treatment of incomplete information). 

Recent articles investigate $\xpathd$ from a modal logic perspective.  
For example, the complexity of the satisfiability and evaluation problems of different fragments are discussed
in~\cite{FigPhD,FigueiraS11,Figueira12ACM,Fig13}, while model theory and
expressivity are studied
in~\cite{ADF14,ICDT14,ICDT14Jair,ADF14journal,KR16,GonzalezA21}.  

In this article, we will focus on reasoning methods for $\xpathd$. 
In this line of work, a Gentzen-style sequent calculus is given for a
very restricted fragment of $\xpathd$, named DataGL, in~\cite{BaeldeLS15}. In DataGL, data
comparisons are allowed only between the evaluation point and its
successors. An extension of the equational axiomatic system
from~\cite{cateLM10} is introduced in~\cite{ADFF16report} that is proved sound and complete for \xpathd. This extension allows
downward navigation and equality/inequality tests.
More recently, \cite{ArecF:hilb16,ArecesF21} provide an axiomatization for this logic,
extended with upward navigation, and with nominals and satisfiability
modalities from Hybrid Logic~\cite{arec:hybr05b}. As argued in, e.g., \cite{Brauner07}, features from Hybrid Logic can be effectively leveraged in the design of proof methods for modal logics.  In~\cite{ArecesFS17},
we introduced sound and complete tableau calculus for hybrid $\xpathd$.

\paragraph{Contributions.}  
In this article we start by introducing a sound, complete and terminating tableau
calculus for $\xpathd$ with downward navigation, where node expressions
are extended with nominals (special labels that are true in only one
node), and path expressions are extended with the satisfiability modality
(allowing the navigation to some particular named node), over the class of all data-graphs. We call this
logic $\Hxpath$. 

Taking inspiration from~\cite{BolanderB07}, we use the hybrid logic operators to provide internalized tableau rules, which we prove are sound and complete for \Hxpath.  The main intuition is
that nominals and satisfiability modalities can be used in tableaux to
keep track of the evaluation of an expression during an attempt to build a
model. This enables us to define tableaux without any help of external features, like labels or relational annotations, as they can be internalized into the language of $\Hxpath$. Crucially, we show how the hybrid machinery can also be used to help with data comparisons. 
We show that the tableau calculus can be ran in polynomial space,
without compromising completeness, establishing in this way that the satisfiability problem
for $\Hxpath$ is $\pspace$-complete.

We then extend our tableaux to handle different structures. First, we show how to enforce certain reachability conditions over the accessibility relations. With this at hand, we provide a method to check satisfiability of $\Hxpath$ over the class of data trees, and show that the complexity of the satisfiability problem remains in \pspace. 
Then, we extend the calculus with so-called pure axioms and node creating rules. These two extensions can be used to characterize many other model classes, and introduce new operators. We show that soundness and  completeness of the resulting tableau calculi follow automatically (even though termination can be compromised). 

This work is an extension of~\cite{ArecesFS17} and improves on it in several ways.  Most importantly we provide details on the termination argument, which was only sketched in the previous article, and show that the $\pspace$ upper bound holds also for trees. To this end, we start by introducing changes to the language that simplify proofs. Concretely, we extend the base language with primitive expressions of the form $\tup{\pathh}\varphi$. This enables us to obtain internalized tableaux, and consequently to prove termination  in a simpler way. Second, we reduce the number of rules by exploring only the left side in a data comparison and adding a commutativity rule. Finally, the current presentation includes a detailed algorithm suitable for implementation.

The contributions in this article serve as a starting point for research and mechanization of other data-aware logics. For example our approach can be use to investigate the logical properties of languages like GQL~\cite{FrancisGGLMMMPR23,FrancisGGLMMMPR23b} and SHACL~\cite{Ortiz23,Ahmetaj0S23}, Modal Logics with Concrete Domains (see, e.g.,~\cite{DemriQ21}), and to study connections with Register Automata (see, e.g.,~\cite{KaminskiF94}).


\paragraph{Organization.}  
\Cref{sec:basic} introduces the syntax
and semantics of $\Hxpath$.  \Cref{sec:tableaux} deals with a tableau calculus for
$\Hxpath$, and with examples of how to build tableaux for concrete expressions. This section also shows the calculus is sound and complete with respect to the class of all data models.
\Cref{sec:termination} deals with the proof of termination of the calculus, whereas~\Cref{sec:complexity} shows that the satisfiability problem for $\Hxpath$ is \pspace-complete.
\Cref{sec:extensionplus} extends the calculus to work
over the class of forest and tree models, and discusses completeness and complexity, showing that the complexity of the satisfiability problem over these classes remains \pspace-complete. 
\Cref{sec:pure-ext} introduces extensions of the calculus with pure axioms and node creating rules. These new rules permits the characterization of new classes of models and the definition of new operators. we argue that the resulting tableau calculi remain sound and complete (but termination might be compromised). \Cref{sec:final} includes final remarks and discusses future lines of research.


\section{Preliminaries}
\label{sec:basic}

We begin by introducing the syntax and the semantics of $\Hxpath$.
We assume a fixed \emph{signature} of pairwise disjoint sets: $\props = \set{p, q, r, \dots}$ of \emph{propositional} symbols, $\noms = \set{i, j, k, \dots}$ of \emph{nominals}, $\rels = \set{\astep[a], \astep[b], \astep[c], \dots }$ of symbols for \emph{accessibility relations}, and $\cmps = \set{\cmp, \mathsf{f}, \mathsf{g},\dots}$ of \emph{comparisons by data (in)equality}.
We further assume that $\props$ and $\noms$ are countably infinite, and that $\rels$ and $\cmps$ are finite.
    
\begin{defi}\label{def:language}
  The \emph{language} of $\Hxpath$ consists of \emph{path} expressions $\alpha$, $\beta$, $\gamma$, \dots, and \emph{node} expressions $\varphi$, $\psi$, $\chi$, \dots.
  Path and node expressions are defined by mutual recursion according to the grammar:
  \begin{align*}
    \alpha, \beta & \Coloneqq
        \pathh \mid
        \goto{i} \mid
        \varphi? \mid
        \alpha \beta \mid
        \alpha \cup \beta \\
    \varphi, \psi & \Coloneqq
        p \mid
        i \mid
        \lnot \varphi \mid
        \varphi \land \psi \mid
        i{:}\varphi \mid 
        \tup{\child} \varphi \mid
        \tup{\alpha =_{\cmp} \beta} \mid
        \tup{\alpha \neq_{\cmp} \beta},
  \end{align*}
  \noindent where $p\in\props$, $i\in\noms$, $\pathh\in\rels$, and $\cmp\in\cmps$.
  By way of notation, we use~$\cmpr$ when there is no need to distinguish between $=_\cmp$ and $\neq_\cmp$.
  We sometimes refer to node expressions of the form $i{:}\varphi$, $\tup{@_i\alpha \cmpr @_j\beta}$, or $\lnot\tup{@_i\alpha \cmpr @_j\beta}$, as \emph{prefixed expressions}. 
\end{defi}


In the syntax above we find nominals and satisfiability statements. The latter come in two flavours: as path expressions of the form $@_i\alpha$, and as node expressions of the form $i{:}\varphi$. The modal diamond $\tup{\child}\varphi$ is included as primitive (as we will see, this has a positive impact in the definition of tableaux in~\Cref{sec:tableaux}).  Notice that we are in a multi-modal system, as we have one modal diamond for each $\pathh\in\rels$. We also allow for different (in)equality comparisons, one for each $\cmp\in\cmps$. 
Finally, we make use the following abbreviations for node expressions:
\begin{align*}
  \varphi \to \psi & \coloneqq \lnot\varphi \lor \psi   &
  \varphi \liff \psi & \coloneqq (\varphi \to \psi) \land (\psi \to \varphi) &
  \varphi \lor \psi & \coloneqq \lnot(\lnot\varphi \land \lnot\psi)  \\
  \tup{\alpha}\varphi & \coloneqq \tup{\alpha\varphi? = \alpha\varphi?} &
  [\alpha]\varphi & \coloneqq \lnot\tup{\alpha}\lnot\varphi &
  [\alpha \cmpr \beta] & \coloneqq \lnot\tup{\alpha \cmpd \beta}.
\end{align*}
The symbol ${\cmpd}$ above is ${\neq_\cmp}$ if ${\cmpr}$ is ${=_\cmp}$, and is ${=_\cmp}$ if ${\cmpr}$ is ${\neq_\cmp}$.


The semantics of path and node expressions is given with respect to \emph{hybrid data models}.

\begin{defi}\label{def:models}
  A \emph{(hybrid data) model} is a tuple $\model = \tup{N, \set{R_\pathh}_{\pathh \in \rels}, \set{\approx_{\cmp}}_{\cmp\in\cmps}, \valuation, \naming}$ where:
  \begin{enumerate} 
    \item $N$ is a non-empty set of elements called \emph{nodes}, 
    \item each $R_{\pathh}\subseteq {N\times N}$ is a binary \emph{accessibility} relation,
    \item each ${\approx_{\cmp}} \subseteq {N\times N}$ is an \emph{equivalence} relation, 
    \item $\valuation :\props \to 2^N$  is a \emph{valuation} function, and 
    \item $\naming : \noms \to N$ is a \emph{naming} function.
  \end{enumerate}
\end{defi}

In brief, the models of $\Hxpath$ can be understood as the typical models of Hybrid Modal Logic, see, e.g.,~\cite{arec:hybr05b}, augmented with a collection of equivalence relations intended to capture equality criteria between data values in nodes.
The \emph{satisfiability} relation is defined below. Notice how satisfiability of a path expression on a model is relative to two nodes (as it describes the accessibility from one to the other), whereas satisfiability of a node expression is at a particular node (stating truth at that node). 

Precisely, the path expression $\pathh$ can be understood as traversing an edge labelled by $\pathh$ in the graph.
In turn, $\goto{i}$ can be understood as a ``go to'' statement, i.e., jump to \emph{the} node named by the nominal~$i$.
Then, $\varphi?$ can be understood as a ``test on the truth of $\varphi$''. 
Finally, $\alpha\beta$ can be understood as the ``concatenation'' of $\alpha$ and $\beta$, 
and $\alpha \cup \beta$ as the ``non-deterministic choice'' between $\alpha$ and $\beta$. 
The intuitive semantics of the node expressions of the form $p$, $i$, $\lnot\varphi$, $\varphi \land \psi$, $i{:}\varphi$, and $\tup{\pathh}\varphi$ are the same as in Hybrid Modal Logic (see, e.g.,~\cite{arec:hybr05b} for details). 
Specifically, 
$i{:}\varphi$ can be understood as ``$\varphi$ is true at the node named by $i$'', while $\tup{\pathh}\varphi$ is read as ``$\varphi$ is true after traversing some  $\pathh$-edge''.
Finally, the intuitive semantics of data equalities $\tup{\alpha =_{\cmp} \beta}$ (and data inequalities $\tup{\alpha \neq_{\cmp} \beta}$) is ``there are $\alpha$ and $\beta$ paths with a common starting point whose end points have data values that are (not) equal according to the criteria determined by $\cmp$''.

\begin{defi}\label{def:semantics}
  Let $\model = \tup{N, \set{R_\pathh}_{\pathh \in \rels}, \set{\approx_{\cmp}}_{\cmp\in\cmps}, \valuation, \naming}$ be a model, and $n,n'\in N$. Then: 
  \[
    \begin{array}{l@{\quad}l@{\quad}l}
      \model, n, n' \vDash \child
        & \text{\it iff} &   n R_{\child} n'   \\
      \model, n, n' \vDash \goto{i}
        & \text{\it iff } &  \naming(i) = n' \\ 
      \model, n, n' \vDash  \varphi?
        & \text{\it iff} & n=n' \text{ and } \model, n \vDash   \varphi  \\
      \model, n, n' \vDash \alpha\beta
        & \text{\it iff} & \text{exists } z \text{ s.t.\ } \model,  n, z \vDash \alpha \text{ and }  \model,  z, n' \vDash \beta \\
      \model, n, n' \vDash \alpha \cup \beta
        & \text{\it iff} & \model, n, n' \vDash  \alpha  \text{ or } \model, n, n' \vDash  \beta \\
      \model, n    \vDash p
        & \text{\it iff} &  n \in \valuation(p)\\
      \model, n    \vDash i
        & \text{\it iff} &  \naming(i) = n \\
      \model, n    \vDash \lnot \varphi
        & \text{\it iff} &   \model, n \nvDash  \varphi \\
      \model, n    \vDash \varphi \land \psi
        & \text{\it iff} &   \model, n \vDash  \varphi \text{ and }  \model, n \vDash  \psi  \\
      \model, n    \vDash  \tup{\pathh}\varphi
        & \text{\it iff} & \text{exists } z \text{ s.t.\ } nR_{\pathh}z \text{ and } \model, z \vDash \varphi \\
      \model, n   \vDash  i{:}\varphi
        & \text{\it iff } & \model, \naming(i) \vDash \varphi \\
      \model, n    \vDash \tup{\alpha =_\cmp \beta}
        & \text{\it iff} & \text{exist } z, z' \text{ s.t.\ } \model, n, z \vDash  \alpha \text{, } \model, n, z' \vDash  \beta \text{, and } z \approx_\cmp z' \\
      \model, n    \vDash \tup{\alpha \neq_\cmp \beta}
        & \text{\it iff} & \text{exist } z, z' \text{ s.t.\ } \model, n, z \vDash  \alpha \text{, } \model, n, z' \vDash  \beta \text{, and } z \not\approx_\cmp z'.
    \end{array}
  \]
  A node expression $\varphi$ is satisfiable iff there is $\model, n$ s.t., $\model, n \vDash \varphi$.
  In turn, $\varphi$ is \emph{valid}, written $\vDash \varphi$ iff $\model, n \vDash\varphi$, for all $\model, n$.
  These notions are analogously defined for path expressions. 
\end{defi}

We conclude this section by listing some immediate properties for node expressions.
These properties are useful in the justification of tableaux rules in \Cref{def:tableau}.

\begin{prop}\label{prop:equivalences}
  Let $\model$ be any model. The following node expressions are valid in $\model$:
  \begin{align*}
    i{:}\tup{\alpha \cmpr \beta}
      & \liff \tup{@_i\alpha \cmpr @_i\beta} &
    i{:}\lnot\tup{\alpha \cmpr \beta}
      & \liff \lnot\tup{@_i\alpha \cmpr @_i\beta} \\
    \tup{\varphi?\alpha \cmpr \beta}
      & \liff \varphi \land \tup{\alpha \cmpr \beta} &
    \lnot\tup{\varphi?\alpha \cmpr \beta}
      & \liff \lnot\varphi \lor \lnot\tup{\alpha \cmpr \beta} \\
    \tup{(\alpha \cup \beta)\gamma \cmpr \eta}
      & \liff \tup{\alpha\gamma \cmpr \eta} \lor \tup{\beta\gamma \cmpr \eta} &
    \lnot\tup{(\alpha \cup \beta)\gamma \cmpr \eta} & \liff \lnot\tup{\alpha\gamma \cmpr \eta} \land \lnot\tup{\beta\gamma \cmpr \eta}.
  \end{align*}
\end{prop} 
\section{Tableaux for $\Hxpath$} \label{sec:tableaux}

In this section we present the notion of tableaux for $\Hxpath$, comment on the view of these tableaux as a calculus for satisfiability, and prove the soundness and completeness of this calculus.
Our ideas and results build on the terminating tableau calculus for Hybrid Modal Logic in~\cite{BolanderB07}.
In brief, we incorporate to such a calculus rules for handling data (in)equalities $\tup{\alpha \cmpr \beta}$ and $[\alpha \cmpr \beta]$.
We consider our presentation of tableaux for $\Hxpath$ simpler and more elegant than the one presented in~\cite{ArecesFS17}.
First, the internalization of rules in our calculus simplifies their manipulation.
Second, our way of handling the left-hand-side of data (in)equalities together with a commutation rule significantly reduces the number of rules.
For the remainder of the article, we assume that $\noms$ is the set of natural numbers.

\subsection{Tableau Calculus.}
\label{subsec:rules}

We begin with the definition of a tableau for a node expression.
We assume familiarity with the usual terminology on labelled trees.
In particular, the  notion of a root, a node, a successor of a node, a branch, a leaf, and the label of a node.
Moreover, we assume some familiarity with tableaux for Hybrid Logic (see~\cite{BolanderB07}).

\begin{defi}\label{def:tableau}
    Let $\varphi$ be a node expression and $i$ a new nominal.
    A \emph{tableau} for $\varphi$ is a labelled tree constructed from an initial single node tree labelled with $i{:}\varphi$ using the branch \emph{extension} rules in \Cref{rules:basic,rules:nominals,rules:internalization,rules:path}. Crucially, rules can be applied only once to its premisses. 
    In this definition, a \emph{new} nominal is the smallest nominal that is larger than all other nominals appearing in the tableau (thus far constructed).
    In turn, a \emph{root} nominal is one appearing in $\varphi$.
    Finally, a node expression $i{:}\tup{\pathh}j$ is an \emph{accessibility constraint} iff it appears in the conclusion of the rules ($\Diamond$) and (child).
\end{defi}


\begin{figure}[p]
    \centering
    \small
    \begin{tabular}{ccccc}
        \toprule
        \multicolumn{5}{c}{\sc Basic}
        \tabularnewline
        \midrule
        {\infer[(\lnot\lnot)]
            {\deduce[]
                {\phantom{|}}
                {\phantom{|}i{:}\varphi}}
            {\deduce[]
            {\phantom{|}{i{:}\lnot \lnot \varphi}}
            {}
            }      
        }
        &    
        {\infer[(\land)]
            {\deduce[]
                {\phantom{|}i{:}\varphi}
                {\phantom{|}i{:}\psi}}
            {\deduce[]
            {\phantom{|}i{:}(\varphi \land \psi)}
            {}
            }
        }
        &
        {\infer[(\lnot \land)]
            {\deduce[]
            {\phantom{|}}
            {\phantom{|}{i{:}\lnot\varphi} \quad {i{:}\lnot\psi}}
            }
            {\deduce[]
            {\phantom{|}{i{:}\lnot(\varphi \land \psi)}}
            {}
            }
        }
        &
        {\infer[(\Diamond^{1})]
            {\deduce[]
            {\phantom{|}{j{:}\varphi}}
            {\phantom{|}{i{:} \tup{\child} j}}
            }
            {\deduce[]
            {\phantom{|}{i{:} \tup{\child} \varphi}}    
            {\phantom{|}}
            }
        }
        &
        {\infer[(\lnot \Diamond^2)]
            {\deduce[]
            {\phantom{|}}
            {\phantom{|}{j{:} \lnot \varphi}}
            }
            {\deduce[]
            {\phantom{|} {i{:} \tup{\child} j}}    
            {\phantom{|}{i{:} \lnot \tup{\child} \varphi}}
            }
        }
        \tabularnewline
        \bottomrule
        \multicolumn{5}{l}{\footnotesize $^1$ $i{:}\tup{\child}\varphi$ is not an accessibility constraint. $j$ is a new nominal.}
        \tabularnewline
        \multicolumn{5}{l}{\footnotesize $^2$ $i{:}\tup{\child}j$ is an accessibility constraint.}
    \end{tabular}
    \caption{Basic Rules}
    \label{rules:basic}
\end{figure}

\begin{figure}[p]
    \centering
    \small
    \begin{tabular}{cccccc}
        \toprule
        \multicolumn{6}{c}{\sc Satisfiability and Nominals}
        \tabularnewline
        \midrule
        {\infer[(\mathrm{nom})]
            {\phantom{|}{j{:}\varphi}}
            {\phantom{|}i{:} (j{:} \varphi)}
        }
        &
        {\infer[(\lnot \mathrm{nom})]
            {\phantom{|}{j{:} \lnot \varphi}}
            {\phantom{|}{i{:} \lnot (j{:} \varphi)}}
        }
        &         
        {\infer[(\mathrm{copy}^1)]
            {\phantom{|} j{:} \varphi }
            {\deduce[]
                {\phantom{|}i{:}j}
                {\phantom{|}i{:}\varphi}
            }
        }
        &
        {\infer[(\mathrm{ref}^2)]
            {\phantom{|} i{:}i}
            {\deduce[]
            {\phantom{|}}           
            {\deduce[]
            {\phantom{|}}
            {\phantom{|}} 
            }
            }
        }                          
        &
        {\infer[(\mathrm{sym})]
            {\phantom{|} i{:}j }
            {\deduce[]
            {\phantom{|}j{:}i}
            {\phantom{|}} 
            }
        }
        &
        {\infer[(\mathrm{trans})]
            {\phantom{|} j{:}k }
            {\deduce[]
            {\phantom{|}j{:}l}
            {\deduce[]
            {\phantom{|} i{:}l} 
            {\phantom{|} i{:}k} 
            }
            }                        
        }
        \tabularnewline
        \bottomrule
        \multicolumn{6}{l}{\footnotesize $^1$ $\varphi$ is not a nominal, $j$ is a root nominal, $j<i$.}
        \tabularnewline
        \multicolumn{6}{l}{\footnotesize $^2$ $i$ is a nominal in the branch.}
    \end{tabular} 
    \caption{Rules for Nominals}
    \label{rules:nominals}
\end{figure}

\begin{figure}[p]
    \centering
    \small
    \begin{tabular}{cccc}
        \toprule
        \multicolumn{4}{c}{\sc Internalization into Data (In)Equalities}
        \tabularnewline
        \midrule
        {\infer[(\mathrm{int}_1)]
            {{\phantom{|}}\langle \goto{i}\alpha \cmpr \goto{i}\beta \rangle}
            {i{:} \langle \alpha \cmpr \beta \rangle}
        }
        &
        {\infer[(\mathrm{int}_2)]
            {\phantom{|} \lnot \langle \goto{i}\alpha \cmpr \goto{i}\beta \rangle }
            {\phantom{|}i{:} \lnot \langle \alpha \cmpr \beta \rangle}
        }
        &
        {\infer[({\lnot}{\cmpr})]
            {\phantom{|}\langle \goto{i} \cmpd \goto{j} \rangle}
            {\phantom{|} \lnot \langle \goto{i} \cmpr \goto{j} \rangle} 
        }
        &
        {\infer[(\mathrm{dRef})]
            {\phantom{|} \langle \goto{i} =_\cmp \goto{j} \rangle }
            {\phantom{|} i{:}j}
        }
        \tabularnewline
        \bottomrule
    \end{tabular} 
    \caption{Rules for Internalizing Nominals and Negations into Data (In)Equalities}
    \label{rules:internalization}
\end{figure}

\begin{figure}[p]
    \centering
    \small
    \begin{tabular}{c@{\quad}c@{\quad}c}
        \toprule
        \multicolumn{3}{c}{\sc Path Expressions in Data (In)Equalities}
        \tabularnewline
        \midrule
        {\infer[(\mathrm{child}^1)]
            {\deduce[]      
            {\phantom{|}\langle \goto{j} \alpha \cmpr \beta \rangle } 
            {\phantom{|} i{:} \langle \child \rangle j  }     
            }    
            {\deduce[]
            {\phantom{|}\langle \goto{i} \child \alpha \cmpr \beta \rangle}
            {\phantom{|}} 
            }
        }
        &
        {\infer[(\lnot \mathrm{child})]
            {\deduce[]      
            {\phantom{|}} 
            {\phantom{|}  \lnot \langle \goto{j} \alpha \cmpr \beta \rangle  }     
            }    
            {\deduce[]
            {\phantom{|}i{:} \langle \child \rangle j }
            {\phantom{|}\lnot \langle \goto{i} \child \alpha \cmpr \beta \rangle} 
            }
        }
        &
        {\infer[(\mathrm{com}_1)]
            {\deduce[]
            {\phantom{|}}
            {\langle \beta \cmpr \goto{i} \rangle}}
            {\langle \goto{i} \cmpr \beta \rangle } 
        }
        \tabularnewline
        {\infer[(\mathrm{test})]
            {\deduce[]
            {\phantom{|}  \langle \goto{i} \alpha \cmpr \beta \rangle}
            {\phantom{|} i{:}\varphi }      
            }    
            {\deduce[]
            {\phantom{|} \langle \goto{i} \varphi? \alpha \cmpr \beta \rangle} 
            {\phantom{|} } 
            }  
        }
        &
        {\infer[(\lnot \mathrm{test} )]
            {\deduce[]
            {\phantom{|} }
            {\phantom{|}  i{:} \lnot \varphi \quad \lnot \langle \goto{i} \alpha \cmpr \beta \rangle }      
            }    
            {\deduce[]
            {\phantom{|} \lnot \langle \goto{i} \varphi? \alpha \cmpr \beta \rangle} 
            {\phantom{|} } 
            }  
        }
        &
        {\infer[(\mathrm{com}_2)]
            {\deduce[]
            {\phantom{|}}
            {\lnot\langle \beta \cmpr \goto{i} \rangle}}
            {\lnot\langle \goto{i} \cmpr \beta \rangle } 
        }
        \tabularnewline
        {\infer[(@)]
            {\phantom{|}  \langle \goto{j} \alpha  \cmpr \beta \rangle }
            {\deduce[]
            {\phantom{|} \langle \goto{i} \goto{j} \alpha \cmpr \beta \rangle} 
            {\phantom{|}} 
            }  
        }
        &
        {\infer[(\lnot @ )]
            {\phantom{|} \lnot \langle \goto{j} \alpha  \cmpr \beta \rangle }
            {\deduce[]
            {\phantom{|} \lnot \langle \goto{i} \goto{j} \alpha \cmpr \beta \rangle} 
            {\phantom{|}} 
            }  
        }
        &
        {\infer[(\mathrm{dTrans})]
            {\langle \goto{i} =_\cmp \goto{j}  \rangle}
            {\deduce[]
            { \langle \goto{k} =_\cmp \goto{j}  \rangle }
            {\langle \goto{i} =_\cmp \goto{k}  \rangle}}
        }
        \tabularnewline
        {\infer[(\cup)]
            {\deduce[]
            {\phantom{|}  }
            {\phantom{|} \langle \goto{i} \alpha\gamma \cmpr \eta \rangle \quad \langle \goto{i} \beta\gamma \cmpr \eta \rangle }      
            }    
            {\deduce[]
            {\langle \goto{i} (\alpha \cup \beta) \gamma \cmpr \eta \rangle} 
            {\phantom{|} } 
            }  
        }
        &
        {\infer[(\lnot \cup )]
            {\deduce[]
            {\phantom{|} \lnot \langle \goto{i} \beta\gamma \cmpr \eta \rangle }      
            {\phantom{|}  \lnot \langle \goto{i} \alpha\gamma \cmpr \eta \rangle }
            }    
            {\deduce[]
            {\phantom{|} \lnot \langle \goto{i} (\alpha \cup \beta) \gamma \cmpr \eta \rangle} 
            {\phantom{|}} 
            }  
        }
        \tabularnewline
        \bottomrule
        \multicolumn{2}{l}{\footnotesize $^1$ $j$ is a new nominal.}
    \end{tabular} 
    \caption{Rules for Handling Path Expressions in Data (In)Equalities}
    \label{rules:path}
\end{figure}

Let us briefly discuss how tableaux rules are meant to be understood. 
In general, all rules are taken to be read as: if the premisses of the rule are in the branch, then, extend the branch by adding to its leaf node successors labelled with the conclusions of the rule. Each rule is applied only once to a given set of premisses present in a branch.
\Cref{rules:basic} presents the usual rules for Boolean and modal operators of Basic Modal Logic, with the distinguishing characteristic of accessibility constraints being internalized in the language using nominals.
\Cref{rules:nominals} presents standard rules for handling nominals, similar to those found in~\cite{BolanderB07,Brauner2011}.
The only difference is the rule (copy) whose side conditions are in place to guarantee termination of the calculus. 
\Cref{rules:internalization} introduces rules for data comparison, they internalize nominals into data (in)equalities.
More precisely, notice how the rule ($\mathrm{int}_1$) takes a satisfiability statement $i{:}\tup{\alpha \cmpr \beta}$, and internalizes $i{:}$ as an $@_i$-prefix in $\alpha$ and $\beta$ resulting in $\tup{@_i\alpha \cmpr @_i\beta}$. In doing this, we anchor the point of evaluation of the node expression at the beginning of each path in the data comparison.
This enables us to work out paths individually using the rules in~\Cref{rules:path}. The rule ($\mathrm{int}_2$) does the same job in presence of a negation. In turn, ($\neg{\cmpr}$) internalizes negated equalities (inequalities) into inequalities (equalities). 
Finally, (dRef) tells us that nodes with the same name must have the same data values.
\Cref{rules:path} depicts the last set of rules in our tableaux.
These rules tell us how to deal with path expressions in data comparisons. 
Essentially, (child) mimics what is done in ($\Diamond$) but in the context of a data comparison, whereas ($\lnot$child) is analogous to (${\lnot}{\Diamond}$).
The rule (test) eliminates $\varphi$ from the path expression and evaluates it as a standalone node expression, while the case for ($\lnot$test) is dual. 
The rules ($@$) and (${\lnot}{@}$) can be understood in analogy with (nom) and (${\lnot}$nom).
The rationale behind ($\cup$) and (${\lnot}{\cup}$) is immediately understood from \Cref{prop:equivalences}. 
A novelty of our calculus is that rules only decompose, step-by-step, path expressions on the left side of a data comparison.
The treatment of the right side takes over after applying ($\mathrm{com}_1$) or  ($\mathrm{com}_2$). 
Finally, (dTrans) corresponds to transitivity of equalities.

\begin{exa}\label{ex:usetableaux}
\Cref{fig:atableau} depicts a tableau for $\tup{\child}\tup{@_2\astep[b]2? =_\cmp \astep[b](q\land 3)?}$, starting from an initial node labelled by $4{:}\tup{\child}\tup{@_2\astep[b]2? =_\cmp \astep[b](q\land 3)?}$.
The annotations on the right indicate the rules we applied and the nodes we applied them to.
The tableau has a single branch, which becomes saturated if we apply the rules (ref) and (dRef) to all the nominals it contains --e.g., if we apply (ref) to $2$, we obtain a node 18 with label $2{:}2$, and if we apply (dRef) to the new node, we obtain a node 19 with label $\tup{@_2=_\cmp @_2}$; and so on for all the nominals in the branch.
\begin{figure}[t]
    \begin{center}
    \begin{prooftree}{}%
    [
        4{:}\tup{\child}\tup{@_2\astep[b]2? =_\cmp \astep[b](q\land 3)?},
        just={root}
        [
            4{:}\tup{\child}5, 
            just={($\Diamond$):!u}
            [
                5{:}\tup{ @_2\astep[b]2? =_\cmp \astep[b](q \land 3)?},
                just={($\Diamond$):!uu}
                [
                    \tup{@_5@_2\astep[b]2? =_\cmp @_5 \astep[b](q \land 3)?},
                    just={(int$_1$):!u}
                    [
                        \tup{@_2\astep[b]2? =_\cmp @_5 \astep[b](q \land 3)?},
                        just={(${@}$):!u}
                        [
                            2{:}\tup{\astep[b]} 6,
                            just={(child):!u}   
                            [   
                                \tup{@_62? =_\cmp @_5 \astep[b](q \land 3)?},
                                  just={(child):!u}
                                [
                                    6{:}2, 
                                    just={(test):!u}
                                    [
                                        2{:}6,
                                        just={(sym):!u}
                                        [
                                            \tup{@_6 =_\cmp @_5 \astep[b](q \land 3)?} ,
                                            just={(${\text{test}}$):!uuu}  
                                            [
                                                \tup{ @_5 \astep[b](q \land 3)? =_\cmp @_6} ,
                                                just={(com$_1$):!u} 
                                                [
                                                    5{:}\tup{\astep[b]}7, 
                                                    just={(child):!u}
                                                    [
                                                        \tup{ @_7 (q \land 3)? =_\cmp @_6},
                                                        just={(child):!uu}
                                                        [
                                                            7{:}(q \land 3), 
                                                            just={(test):!u}
                                                            [
                                                                \tup{ @_7 =_\cmp @_6},
                                                                just={(test):!uu}
                                                                [
                                                                    \tup{ @_6 =_\cmp @_7},
                                                                    just={(com$_1$):!u}
                                                                    [
                                                                        7{:}q, 
                                                                        just={($\land$):!uuu}
                                                                        [
                                                                            7{:}3, 
                                                                            just={($\land$):!uuuu}
                                                                            [
                                                                                3{:}7, 
                                                                                just={(sym):!u}
                                                                                [
                                                                                    3{:}q, 
                                                                                    just={(copy):!uuu,!uu}
                                                                                ]
                                                                            ]
                                                                        ]
                                                                    ]
                                                                ]
                                                            ]
                                                        ]
                                                    ]
                                                ]
                                            ]
                                        ]
                                    ]
                                ]
                            ]
                        ]
                    ]
                ]
            ]
        ]
    ]
    \end{prooftree}
\end{center}
   \caption{A tableau for $\tup{\child}\tup{@_2\astep[b]2? =_\cmp \astep[b](q\land 3)?}$}\label{fig:atableau}
\end{figure}
\end{exa}

We consider our tableaux to be simpler and more elegant than the proposal  in~\cite{ArecesFS17}. Treating only the left side of data comparisons significantly reduces  the number of rules. It is also clear that our tableaux are \emph{internalized}; they have (only) node expressions as labels. Internalized tableaux have important advantages. In particular, it will enable us to define extended calculi to handle many different frame conditions, and obtain completeness automatically as it is done in~\cite{Blackburn00,Blackburn2002BeyondPA} (see~\Cref{sec:pure-ext}). 
Our tableaux are also \emph{analytic} in the sense of~\Cref{def:analytic} below (see, e.g.,~\cite{Smullyan1968} for background on analytic tableaux). 
This property is helpful to provide a termination argument (see \Cref{sec:termination}).

\begin{defi} \label{def:subformulas}
    The \emph{subcomponents} of a node expression $\varphi$ are defined as
        \begin{longtable}{r@{\ }c@{\ }l}
            $\sub(p)$
            & $=$ & $\set{p}$ \\
            $\sub(i)$
            & $=$ & $\set{i}$ \\
            $\sub(\lnot \varphi)$
            & $=$ & $\set{\lnot \varphi} \cup \sub(\varphi)$ \\
            $\sub(\varphi \land \psi)$
            & $=$ & $\set{\varphi \land \psi} \cup \sub(\varphi) \cup \sub({\psi})$  \\            
            $\sub(i{:}\varphi)$
            & $=$ & $\set{i,\, i{:}\varphi} \cup \sub(\varphi)$ \\
            $\sub(\tup{\pathh} \varphi)$ 
            & $=$ & $\set{ \tup{\pathh} \varphi} \cup \sub(\varphi)$ \\  
            $\sub(\tup{ \patha \cmpr \patha[b] })$
            & $=$ & $\set{\tup{ \patha \cmpr \patha[b]}} \cup \sub'(\patha) \cup \sub'(\patha[b])$ \\
            $\sub(\tup{\patha \cmpr \patha[b]\beta})$
            & $=$ & $\set{\tup{\patha \cmpr \patha[b]\beta}} \cup \sub(\tup{\patha[b]\beta \cmpr \patha})$ \\
            $\sub(\tup{\patha\alpha \cmpr \beta})$
            & $=$ & $\set{\tup{\patha\alpha \cmpr \beta}} \cup \sub(\tup{\alpha \cmpr \beta}) \cup \sub'(\patha)$ \\
            $\sub(\tup{ (\alpha \cup \beta) \gamma \cmpr \xi })$
            & $=$ & $\set{\tup{ (\alpha \cup \beta) \gamma \cmpr \xi }} \cup \sub{(\tup{ \alpha \gamma \cmpr \xi })} \cup \sub{(\tup{ \beta \gamma \cmpr \xi })}$.
        \end{longtable}
    where $\patha \in \{\goto{i}, \pathh, \varphi?\}$. 
    $\sub'(\goto{i}) = \set{i}$, $\sub'(\child) = \emptyset$, and $\sub'(\varphi?) = \sub(\varphi)$.
\end{defi}

Subcomponents are the basic parts tableau rules operate on as they break down complex expressions.
During this process, the rules often prefix them with nominals, associating parts of expressions with specific points in a model. The definition of quasicomponents below formalizes which combinations of nominals and subcomponents can arise during this process.

\begin{defi} \label{def:quasi:component}
        The set of quasi subcomponents of a node expression
        $i{:}\varphi$ is defined as
    \begin{align*}
        \qsub(i{:}\varphi) = \ & \set{j{:}\psi, j{:}\lnot\psi \mid \psi\in\sub(i{:}\psi), j \in \noms}   \\
            &  \cup \{\tup{\goto{j}\alpha \cmpr \goto{k}\beta}, 
            \tup{\goto{j} \cmpr \goto{k}\beta}, 
            \tup{\goto{j}\alpha \cmpr \goto{k}}, 
            \tup{\goto{j} \cmpr \goto{k}} \\ 
            & \quad \ \mid \tup{\alpha \cmpr \beta} \in \sub(i{:}\varphi), j \in \noms\} \\
        & \cup \{\lnot \tup{\goto{j}\alpha \cmpr \goto{k}\beta},
        \lnot \tup{\goto{j} \cmpr \goto{k}\beta},
        \lnot \tup{\goto{j}\alpha \cmpr \goto{k}},
        \lnot \tup{\goto{j} \cmpr \goto{k}},
        \tup{\goto{j} \cmpd \goto{k}}  \\ 
        & 
        \quad \ \mid  \lnot \tup{\alpha \cmpr \beta} \in \sub(i{:}\varphi), j \in \noms\}.
    \end{align*}
\end{defi}

The cases where the quasi subcomponent are $j{:}\psi$, or $j{:}\lnot\psi$ 
arise naturally from the basic tableau rules. 
The other cases, correspond to applications of commutativity rules and for handling data equalities.
For instance, $\tup{\goto{j}\alpha \cmpr \goto{k}\beta}$ and $\lnot\tup{\goto{j}\alpha \cmpr \goto{k}\beta}$ may arise as the result of applying one of the internalization rules in \Cref{rules:internalization}.
The case $\tup{\goto{j} \cmpr \goto{k}\beta}$ corresponds to a situation in which the left-hand side of a data (in)equality has been fully decomposed. 
Similarly, $\tup{\goto{j} \cmpr \goto{k}}$ corresponds to a situation in which both sides of a data equality have been fully decomposed, and
 $\tup{\goto{j} \cmpd \goto{k}}$ may arise from applying (${\lnot}{\cmpr}$).

\begin{defi} \label{def:analytic}
    A tableau for a node expression $\varphi$ is \emph{analytic} iff the label of any node in the tableau is one of the following:
    \begin{enumerate}
        \item a quasi subcomponent of the root,
        \item an accessibility constraint,
        \item an equality constraint $\tup{@_i =_\cmp @_j}$ resulting from the rules (dRef) and (dTrans), or
        \item a satisfiability constraint $i{:}j$ resulting from the rules (ref), (sym), or (trans).
    \end{enumerate}
\end{defi}

\begin{lem} \label{prop:analytic}
    The tableaux in \Cref{def:tableau} are analytic.
\end{lem}
\begin{proof}
    Follows immediately from the tableaux rules.
\end{proof}

\Cref{def:consistency} is auxiliary to seeing tableaux rules as a calculus for satisfiability. 

\begin{defi} \label{def:clash}\label{def:saturated}\label{def:consistency}
    A branch $\Theta$ of a tableau has a \emph{clash} iff one of the following conditions hold:
        \begin{enumerate}
            \item $\{i{:}\varphi,  i{:}\neg \varphi\}\subseteq\Theta$,
            \item $\{\tup{@_i =_\cmp @_j},\tup{@_i \neq_\cmp @_j}\}\subseteq\Theta$.
    \end{enumerate}
    We say that $\Theta$ is \emph{closed} if it contains a clash, otherwise it is \emph{open}.
    The tableau is \emph{closed} iff all its branches are closed, otherwise it is \emph{open}. 
    Finally, we say that $\Theta$ is \emph{saturated} iff it is fully expanded according to the rules of the tableau system.
\end{defi}

The notion of clash tells us that a contradiction has been found in a branch of the tableau.
At the same time, the notion of an open and saturated branch tells us that such a branch is free of contradictions (i.e., contradictions are impossible in the branch).  
To check whether  a node expression $\varphi$ is satisfiable, 
we begin with a single node tableaux having $i{:}\varphi$ as its root and successively apply the rules in~\Cref{rules:basic,rules:nominals,rules:internalization,rules:path}.
We answer the question negatively if at some stage the tableau closes, and answer the question positively if we find an open and saturated branch. 
The correctness of~\Cref{def:consistency} as a calculus for satisfiability follows from the results in~\Cref{sec:soundness,sec:completeness}.
Moreover, the results in~\Cref{sec:termination} tell us that this calculus yields a decision procedure for the satisfiability problem.
Even further, the results in~\Cref{sec:complexity} tells us that the computational complexity of this calculus is relatively low, insofar as it is in the same class as the satisfiability problem for Basic Modal Logic~\cite{handbook06}.
\subsection{Soundness}\label{sec:soundness}

In \Cref{thm:soundness} we present the proof of soundness for tableaux as a calculus for satisfiability.
Our proof follows the same strategy used in~\cite{Fit83,Fit96,Priest08}. 
We start by introducing a well-known satisfiability preserving result regarding the use of new nominals.




\begin{lem}
	\label{lemma:nom}
	\label{lemma:child}
        Let $\varphi$ be a node expression and let $j$ be a nominal not in $\varphi$.
		Then, (1) $\varphi$ is satisfiable iff $j{:}\varphi$ is satisfiable,
        (2) $\varphi = \psi \land i{:}\tup{\pathh}\chi$ is satisfiable iff $\varphi \land i{:}\tup{\pathh} j \land j{:}\chi$ is satisfiable, and 
        (3) $\varphi = \psi \land \tup{@_i \pathh\alpha \cmpr \beta}$ is satisfiable iff $\varphi \land i{:}\tup{\pathh} j \land (\tup{@_j \alpha \cmpr \beta})$ is satisfiable.
\end{lem}

\begin{proof}
	In all cases, the implication from right to left is direct. 
	For (1), let $\model,n\vDash\varphi$. Since $j$ does not appear in $\varphi$, we can build a new model $\model'$ that is exactly like $\model$ with the exception that in $\model'$ we set $\naming(j)=n$. Thus, $\model',n\vDash j{:}\varphi$. 
	%
	For (2) and (3), similarly, if we choose $j$ not in $\varphi$, we can use the same idea and set $\naming(j)=m$ for $m$ such that $(\naming(i),m) \in R_{\child}$. 
\end{proof}




Now we proceed to prove some properties relating the extension of a branch using tableaux rules, and the satisfiability of the node expressions contained in such a branch.

\begin{defi}
    \label{def:faithful}
    \label{def:branch-sat}
    Let $\Theta$ be branch of a tableau.
    We say that $\Theta$ is \emph{faithful} to a model $\model$ iff every node expression in $\Theta$ is satisfiable in $\model$.
    In turn, $\Theta$ is satisfiable iff there is a model $\model$ such that $\Theta$ is faithful to $\model$.
\end{defi}

The following lemma makes clear the way in which tableau rules work: they only introduce a contradiction (a clash) if such a contradiction is already present in the node expression we are building the tableau for.
Stated otherwise, tableau rules preserve satisfiability. 

\begin{lem}
    \label{lemma:soundness}
    The application of any one of the tableau rules in~\Cref{rules:basic,rules:nominals,rules:internalization,rules:path} to a satisfiable branch $\Theta$ yields another satisfiable branch.
\end{lem}

\begin{proof}
    Let $\Theta$ be a satisfiable branch of tableau,  i.e.,
    $\Theta$ is faithful to a model $\model$.
    The proof is by a case analysis of tableau rules.
    We develop some relevant cases.
    For the rules ($\Diamond$) and (child), the result follows from \Cref{lemma:nom}.
    For the rule (copy), notice that we have $\{i{:}\varphi, i{:}j\} \subseteq \Theta$ and that the extension after the application of (copy) is $\Theta' = \Theta \cup \{j{:}\varphi\}$.
    Moreover, we have $\model \vDash i{:}j$ and $\model \vDash i{:}\varphi$.
    From the first conjunct, we obtain $\naming(i) = \naming(j)$, and from the second $\model, \naming(i) \vDash \varphi$, 
    then $\model \vDash j{:}\varphi$.
    Therefore, $\Theta'$ is satisfiable.
    %
    For the rule (dRef),
    we have $i{:}j \in \Theta$ and that the extension after the application of (dRef) is $\Theta' = \Theta \cup \{\tup{@_i =_{\cmp} @_j}\}$.
    Moreover, we have $\model \vDash i{:}j$, and thus $\naming(i) = \naming(j)$.
    Since $\approx_{\cmp}$ is an equivalence relation, $(\naming(i), \naming(j)) \in {\approx_{\cmp}}$, then $\model \vDash \tup{@_i =_{\cmp} @_j}$.
    Therefore, $\Theta'$ is satisfiable.
    %
    Finally, notice that the cases (test), ($\lnot$test), ($\cup$), (${\lnot}{\cup}$), (int$_1$), and (int$_2$) are immediate from \Cref{prop:equivalences}.
\end{proof} 

The result in \Cref{lemma:soundness} enables us to state and prove the main result of this section.

\begin{thm}[Soundness]
    \label{thm:soundness}
    Let $\varphi$ be a node expression. If $\varphi$ is satisfiable, then there is a saturated branch of a tableau for $\varphi$ that is open.
\end{thm}

\begin{proof}
    Aiming for a contradiction, suppose all saturated tableaux for $\varphi$ are closed, but $\varphi$ is satisfiable. Any tableau for $\varphi$ starts with $i{:}\varphi$, for some $i\in\noms$ not appearing in~$\varphi$.
    From~\Cref{lemma:nom}, we know $i{:}{\varphi}$ is also satisfiable, thus the initial branch $\Theta=\set{i{:}\varphi}$ of the tableaux is satisfiable. 
    From~\Cref{lemma:soundness}, we know that the application of the rules in~\Cref{rules:basic,rules:nominals,rules:internalization,rules:path} always yields a satisfiable branch from $\Theta$.
    Eventually, we would obtain a saturated branch.
    This contradicts the assumption that all saturated tableaux built from $\varphi$ are closed.
\end{proof}


\subsection{Completeness}\label{sec:completeness}

Let us now turn our attention to proving the completeness of tableaux as a calculus for satisfiability.
The main idea behind the proof is to look at an open and saturated branch of tableaux and to build a model from it (cf., e.g., \cite{Fit83,Fit96,Priest08}).
We begin with some preliminary definitions and results.


\begin{defi}
    \label{def:prefixrelation}
    We use $\noms(\Theta)$ for the set of nominals appearing in a branch $\Theta$ of a tableau. 
    If $\Theta$ is saturated, for every $\set{i,j} \subseteq \nominal(\Theta)$ define
    $i \equiv_\Theta j \mbox{ iff } i{:}j \in \Theta$, and let $[i]_{\equiv_{\Theta}} = \{ j \mid i \equiv_\Theta j\}$.
\end{defi}

Intuitively, the relation $\equiv_\Theta$ groups together all the nominals in the branch $\Theta$ representing the same node in the model we want to build.
The following lemma states $\equiv_\Theta$ is an equivalence relation as desired.

\begin{lem}
    \label{lemma:equiv-prefixrelation}
    $\equiv_\Theta$ is an equivalence relation.
\end{lem}
\begin{proof}
Immediate from (ref), (sym) and (trans).
\end{proof}

Let us proceed to introduce key elements in our proof completeness: \emph{nominal urfathers}.

\begin{defi}\label{def:urfather} 
    Let $\Theta$ be a saturated tableau branch of a tableau.
    Define $\urf[i]=\min([i]_{\equiv_{\Theta}})$; and we say that
$\urf[i]$ is the \emph{urfather} of $i$.
    The set $\urfather{}{\Theta} = \setof{\urf[i]}{i \in \noms(\Theta)}$ is the set of urfathers in the branch.
\end{defi}

The following is an important property of nominal urfathers.
Intuitively, as pointed out in \cite{Brauner2011}, this property tells us that non-trivial reasoning involving non-singleton equivalence classes is carried out only in connection with nominals appearing in the root node expression.

\begin{lem}\label{lemma:representative}
    If $\Theta$ is a saturated branch of a tableau and $\urf \neq i$, $\urf$ is a root nominal.
\end{lem}
\begin{proof}
    Let $\urf \neq i$ and $\urf = j$.
    The proof proceeds by cases.
    
    \begin{description}
        \item[($i$ is a root nominal)]
        Recall that a root nominal is a nominal appearing in the root of our tableau. Any new nominal added by the application of a rule from~\Cref{subsec:rules} is greater than any root nominal. Since $j < i$ and $i$ is a root nominal, $j$ must be also a root nominal.

        \item[($i$ is not a root nominal)]
        If $i{:}j$ is the root of the tableau, then, we are done.
        If $i{:}j$ is not the root of the tableau, then, it was introduced by a tableau expansion rule.
        It is clear that $j$ is a root nominal if it was introduced by ($\lnot\lnot$), ($\land$), (nom), and (test).
        Notice that in any of these cases, $j$ is a component of the node expression labelling the root of the tableau.
        Otherwise, $i{:}j$ is introduced into the tableau by (sym) or (trans).
        In these two last cases, it must be that there is a root nominal $k$ such that $\set{i{:}k, k{:}k} \subset \Theta$.
        That $j$ is a root nominal follows from this fact and the minimality of being an urfather.
    \end{description}
\end{proof}

From their definition, it follows that nominal urfathers can be taken as canonical representatives of the equivalence classes in $\noms(\Theta)/{\equiv\Theta}$.
This characteristic makes them key elements in our proof of completeness.
Moreover, as made clear in \Cref{lemma:urfclos}, nominal urfathers collect the information about all members of the equivalence class.



\begin{lem}\label{lemma:urfclos}
    If $\Theta$ is a saturated branch of a tableau and $i{:}\varphi \in \Theta$, then, $\urf{:}\varphi \in \Theta$.
\end{lem}
\begin{proof}
    Suppose that  $i{:}\varphi\in\Theta$ and $\urf=j$.
    If $i = j$, then, we are done.
    If $i \neq j$, from \Cref{lemma:representative}, we know $j$ is a root nominal.
    Moreover, from \Cref{def:prefixrelation},  $i{:}j \in \Theta$, and, from \Cref{def:urfather}, we know $j < i$.
    Since $\Theta$ is saturated it is closed under (copy), $j{:}\varphi\in\Theta$.
    Thus, $\urf:\varphi \in \Theta$.
\end{proof}

The results in \Cref{lemma:representative,lemma:urfclos} pave the way to define the intended model from an open branch.
Intuitively, they tell us that if our goal is to build a model that satisfies a node expression $\varphi$ from an open and saturated branch of a tableau, then, we can restrict our attention to nominal urfathers.
In comparison with what is done in~\cite{ArecesFS17}, the use of urfathers simplifies the definition of the model, and aid in the proof of \Cref{prop:truthlemma}.

\begin{defi}[Extracted Model]\label{def:extmodel}
    Let $\Theta$ be an open and saturated branch of a tableau.
    Define a structure
        $ \model^\Theta =
            \tup{
                N^{\Theta},
                \set{R^{\Theta}_\pathh}_{\pathh \in \rels},
                \set{\approx^{\Theta}_{\cmp}}_{\cmp\in\cmps},
                \valuation^{\Theta},
                \naming^{\Theta}
            }$
    where
    \[
        \begin{array}{l@{\ }c@{\ }l}
            N^{\Theta}              & = & \urfather{}{\Theta} \\
            R^{\Theta}_\pathh       & = & \{(\urf,\urfather{\Theta}{j}) \mid  i{:}\tup{\child}j\in\Theta \} \\
            \approx^{\Theta}_{\cmp} & = & \{(\urf, \urfather{\Theta}{j}) \mid \tup{@_i =_\cmp @_{j}} \in\Theta \}\\            
            \valuation^{\Theta}(p) & = & \{\urf \mid  i{:}p \in \Theta \} \\
            \naming^{\Theta}(i) & = & 
                    \begin{cases}
                        \min(\urfather{}{\Theta}) & \text{if }  i \notin\nominal(\Theta)\\
                        \urfather{\Theta}{i} & \text{if } i\in\nominal(\Theta).
                    \end{cases} 	
        \end{array}
    \]
\end{defi}

\begin{prop}\label{prop:extmodel}
    The structure $\model^{\Theta}$ in \Cref{def:extmodel} is a model of $\Hxpath$.
\end{prop}
\begin{proof}
    The only challenge is proving $\approx^{\Theta}_{\cmp}$ to be an equivalence relation. 
    Reflexivity follows from (ref) and (dRef), symmetry follows from (com$_1$), and transitivity follows from (dTrans).
\end{proof}

We refer to $\model^{\Theta}$ as the \emph{extracted model} for $\Theta$.
This is the main ingredient in the proof of completeness of tableaux as a calculus for satisfiability.
The proof of this result is carried out by induction on the \emph{size} of a node expression. 
We introduce this definition of size below and comment on it immediately after. 

\begin{defi}\label{def:size}
    The function $\size$ is defined by mutual recursion on node and path expressions.
    \[\begin{array}{l@{\,}l@{\qquad}@{\qquad}l@{\,}l}
        \size(i) & = 1 & 
        \size(\child ) & = 1 \\ 
        \size(p) & = 1 & 
        \size(@_i) & = 1 \\ 
        \size(\varphi \land \psi) & = 1 + \size(\varphi) + \size(\psi) &
        \size(\varphi?) & = 1 + \size(\varphi) \\
        \size(\lnot \varphi) & = 1 + \size(\varphi) &
        \size( \alpha \beta ) & = \size(\alpha) + \size(\beta) \\
        \size(\tup{\child} \varphi) & = 1 + \size(\varphi) &
        \size( \alpha \cup \beta ) & = 1 + \size(\alpha) + \size(\beta) \\
        \size(i{:}\varphi) & = 3 + \size(\varphi) \\
        \size(\tup{\alpha \cmpr \beta}) & = 5 + \size(\alpha) + \size(\beta).
    \end{array}\]   
\end{defi}

The definition of $\size$ is necessary because we cannot rely on structural induction over node expressions. This is due to the fact that rule applications may generate expressions not structurally contained in the original input. For example, applying ($\lnot$int) to $i{:}\lnot\tup{\alpha \cmpr \beta}$ introduces $\lnot\tup{@_i\alpha \cmpr @_i\beta}$ in the branch, which is not a part of the original expression.
To handle this, we define a size measure on node and path expressions and carry out induction on that measure instead. The constants $3$ for $i{:}\varphi$ and $5$ for $\tup{\alpha \cmpr \beta}$ are chosen to ensure that rule applications strictly reduce size. This guarantees that the induction is well-founded. For instance, in the case of ($\lnot$int), we have $\size(\lnot\tup{@_i\alpha \cmpr @_i\beta}) < \size(i{:}\lnot\tup{\alpha \cmpr \beta})$.

\begin{lem}\label{prop:truthlemma}
    Let $\Theta$ be an open and saturated branch, and $\model^{\Theta}$ be the model extracted from it. 
    In addition, let $\model^{\Theta} \vDash \varphi$ indicate $\model^{\Theta}, n \vDash \varphi$ for all $N^{\Theta}$.
    Then, $\varphi \in \Theta$ implies $\model^{\Theta} \vDash \varphi$.
\end{lem}
\begin{proof}
    Observe that every $\varphi \in \Theta$ has one of the following forms: 
    \[
        (1)~i{:}\psi \qquad\qquad
        (2)~\tup{@_i \alpha \cmpr @_j\beta } \qquad\qquad
        (3)~\lnot\tup{@_i \alpha \cmpr @_j\beta }
    \]
    The proof is by induction on the size of $\varphi \in \Theta$.
    We focus on some of the cases in the proof.
    \begin{description}
        \item[\textnormal{For the base cases we consider}] \hfill
            \begin{description}
                \item[($\varphi = i{:}j$)] 
                    We need to prove $\model^\Theta \vDash i{:}j$, or equivalently, $\urf[i] = \urf[j]$.
                    The result is immediate from the definition of $\naming^{\Theta}$.
                \item[($\varphi = i{:}\lnot j$)]
                    We need to prove $\model^\Theta \vDash i{:}\lnot j$, or equivalently $\urf \neq \urfather{\Theta}{j}$.
                    The proof is by contradiction.
                    Suppose $\urf = \urf[j]$.
                    By \Cref{def:urfather,def:prefixrelation}, $i{:}j \in \Theta$.
                    This implies that $\Theta$ contains a clash, contradicting that $\Theta$ is open. Thus, $\model^\Theta \vDash i{:}\lnot j$. 
                \item[($\varphi = \tup{@_i =_{\cmp} @_j}$)] 
                    We neet to prove $\model^\Theta \vDash \tup{@_i =_{\cmp} @_j}$, or equivalently that the pair ${(\urfather{\Theta}{i}, \urfather{\Theta}{j})} \in {\approx^{\Theta}_{\cmp}}$.
                    The result is immediate from the definition of ${\approx^{\Theta}_{\cmp}}$.
                \item[($\varphi = \tup{ @_i \neq_{\cmp} @_j}$)] 
                    We need to prove $\model^\Theta \vDash \tup{@_i \neq_{\cmp} @_j}$, or equivalently that the pair ${(\urf, \urfather{\Theta}{j})} \notin {\approx^{\Theta}_{\cmp}}$.
                    The proof is by contradiction.
                    Suppose that the pair ${(\urf, \urfather{\Theta}{j})} \in {\approx^{\Theta}_{\cmp}}$.
                    By definition of ${\approx^{\Theta}_{\cmp}}$, $\tup{@_i =_{\cmp} @_j} \in \Theta$.
                    But this means the branch contains a clash, contradicting that $\Theta$ is open. Thus, $\model^\Theta \vDash \tup{@_i \neq_{\cmp} @_j}$. 
            \end{description}
        \item[\textnormal{The inductive hypothesis (IH) is}] for all $\psi \in \Theta$ such that $\size(\psi) < \size(\varphi)$,  $\model^\Theta \vDash \psi$.
        \item[\textnormal{For the inductive cases we consider}] \hfill
        \begin{description}
            \item[($\varphi = i{:}\lnot\lnot \psi$)]
                We need to prove $\model^{\Theta} \vDash i{:}\lnot\lnot\psi$, or equivalently
                    $\model^{\Theta} \vDash i{:}\psi$.
                From $(\lnot\lnot)$, ${i{:}\psi \in \Theta}$.
                By the IH, $\model^{\Theta} \vDash i{:}\psi$.
            \item[($\varphi = i{:}\lnot (\psi \land \chi)$)]
                We need to prove $\model^{\Theta} \vDash i{:}\lnot(\psi \land \chi)$, or equivalently
                    $\model^{\Theta} \vDash i{:}\lnot\psi$ or
                    $\model^{\Theta} \vDash i{:}\lnot\chi$.
                From ($\lnot\land$), $i{:}\lnot\psi \in \Theta$ or $i{:}\lnot\chi \in \Theta$.
                W.l.o.g., let $i{:}\lnot\psi \in \Theta$. By the IH, $\model^{\Theta} \vDash i{:}\lnot\psi$.
            \item[($\varphi = i{:}\lnot j{:}\psi$)]
                We need to prove $\model^{\Theta} \vDash {i{:}\lnot(j{:}\psi)}$, or equivalently
                    $\model^{\Theta} \vDash {j{:}\lnot\psi}$.
                From ($\lnot$nom), $j{:}\lnot\psi \in \Theta$.
                Then, by the IH, $\model^{\Theta} \vDash j{:}\lnot \psi$.
            \item[($\varphi = i{:}\lnot \tup{\pathh} \psi$)]
                We must prove $\model^{\Theta} \vDash {i{:}\lnot \tup{\pathh}\psi}$, or equivalently,
                for all $(\urf,n) \in R^{\Theta}_{\pathh}$, ${\model^{\Theta}, n \vDash \lnot\psi}$.
                Let $(\urf,n) \in R^{\Theta}_{\pathh}$.
                The definition of $R^{\Theta}_{\pathh}$ tells us that exists $i{:}\tup{\pathh}j \in \Theta$ s.t.\ $\urfather{\Theta}{j} = n$.
                Since $\{i{:}\lnot\tup{\pathh}\psi, i{:}\tup{\pathh}j\} \subseteq \Theta$, using ($\lnot \Diamond$), we get $j{:}\lnot\psi \in\Theta$.
                By IH, $\model^\Theta \vDash j{:}\lnot \psi$, i.e., $\model^\Theta, n \vDash \lnot \psi$.
            \item[($\varphi = i{:}\lnot\tup{\alpha \cmpr \beta}$)]
                We neet to prove $\model^{\Theta} \vDash i{:} \lnot \tup{\alpha \cmpr \beta}$.
                Equivalently, we prove $\model^{\Theta} \vDash {\lnot \tup{@_i\alpha \cmpr @_i\beta}}$.
                From (int$_2$), $\lnot\tup{ @_i \alpha \cmpr @_i \beta} \in \Theta$.
                Then, we have $\size(\lnot\tup{ @_i \alpha \cmpr @_i \beta}) < \size(i{:} \lnot \tup{\alpha \cmpr \beta})$.
                This enables us to apply the IH and get $\model^\Theta \vDash \lnot\tup{@_i \alpha \cmpr @_i \beta}$.
            \item[($\varphi = i{:}\tup{\child}\psi$)] 
                We need to prove $\model^{\Theta} \vDash {i{:}\tup{\child}\psi}$.
                From ($\Diamond$), $\set{i{:}\tup{\child}j, j{:}\psi} \subseteq \Theta$. 
                Using the IH, we can prove $\model^\Theta \vDash i{:}\tup{\pathh}j \land j{:}\psi$.
                \Cref{lemma:child} gives us the desired result.
            \item[($\varphi = \tup{@_i (\alpha \cup \beta) \gamma \cmpr \eta}$)]
                We need to prove that $\model^\Theta \vDash \tup{@_i (\alpha \cup \beta) \gamma \cmpr \eta}$, equivalently $\model^\Theta \vDash \tup{@_i\alpha\gamma \cmpr \eta}$ or $\model^\Theta \vDash \tup{@_i \beta\gamma \cmpr \eta}$.
                From ($\cup$), $\tup{@_i\alpha\gamma \cmpr \eta} \in \Theta$ or $\tup{@_i\beta\gamma \cmpr \eta} \in \Theta$.
                W.l.o.g.\ let $\tup{@_i\alpha\gamma \cmpr \eta} \in \Theta$, by the IH, $\model^{\Theta} \vDash \tup{@_i\alpha\gamma \cmpr \eta}$.
            \item[($\varphi = \lnot\tup{@_i(\alpha \cup \beta) \gamma \cmpr \eta}$)]
                We need to prove  $\model^\Theta \vDash \lnot\tup{@_i (\alpha \cup \beta) \gamma \cmpr \eta}$, equivalently $\model^\Theta \vDash \lnot\tup{@_i\alpha\gamma \cmpr \eta}$ and $\model^\Theta \vDash \lnot\tup{@_i \beta\gamma \cmpr \eta}$.
                From (${\lnot}{\cup}$), $\lnot\tup{@_i\alpha\gamma \cmpr \eta} \in \Theta$ and $\lnot\tup{@_i\beta\gamma \cmpr \eta} \in \Theta$.
                By the IH, $\model^{\Theta} \vDash \lnot\tup{@_i\alpha\gamma \cmpr \eta}$ and $\model^{\Theta} \vDash \tup{@_i\beta\gamma \cmpr \eta}$.
        \end{description}

    \end{description}
\end{proof}

The lemma above is the fundamental piece to state the intended completeness result.

\begin{thm}[Completeness]
    Let $\varphi$ be a node expression.
    If there is a tableau for $\varphi$ with an open and saturated branch, then, $\varphi$ is satisfiable.
\end{thm}
\begin{proof}
    Suppose that there is a tableau for $\varphi$ with an open and saturated branch $\Theta$.
    Let $\model^{\Theta}$ be the extracted model for $\Theta$.
    From \Cref{prop:truthlemma}, we know $\model^{\Theta} \vDash i{:}\varphi$ where $i{:}\varphi$ is the root of the tableau.
    Immediately, $\model^{\Theta}, \naming(i) \vDash \varphi$.
    Thus, $\varphi$ is satisfiable.
\end{proof}

As it is usual, our tableaux-based calculus not only provides a way to check the satisfiability of a node expression, but also acts as a \emph{model building procedure}, since for every satisfiable node expression it is possible to obtain a model (\Cref{def:extmodel}). 
We finish this section by showing how the extracted model is built for a given node expression.

\begin{exa}
    \label{ex:extracted model}
    Consider the tableau for $\varphi = \tup{\child}\tup{@_2\astep[b]2? =_\cmp \astep[b](q\land 3)?}$ in~\Cref{ex:usetableaux}.
    This tableau has a sole open and saturated branch $\Theta$.
    Thus, we can conclude $\varphi$ is satisfiable.
    We depict the extracted model from $\Theta$ in \Cref{ex:extracted:model}.
    In this figure, nodes are labelled by the nominals and propositional symbols satisfied by them.
    We omit the reflexive edges for $\approx_\cmp$.
    \begin{figure}[t]
        \centering
        \begin{tikzpicture}%
            [
                every node/.style={scale=.7},
                modal
            ]
            \node[world] (4) [label=225:{\footnotesize $4$}] {$4$};
            \node[world] (5) [label=225:{\footnotesize $5$},below=of 4] {$5$};
            \node[world] (7) [label=315:{\footnotesize $3$},below=of 5] {$3$, $7$, $q$};
            \node[world] (6) [label=225:{\footnotesize $2$},left=2cm of 7] {$2$, $6$};

            \path[->]
                (4)
                edge
                node[midway, right]{\footnotesize$\mathsf{a}$}
                (5);
            \path[->]
                (5)
                edge
                node[midway, right]{\footnotesize$\mathsf{b}$}
                (7);
            \path[-]
                (7)
                edge[dashed, bend right]
                node[midway, above]{\footnotesize$\approx_\cmp$}
                (6);
            \path[->]
                (6)
                edge[reflexive]
                node[midway, above]{\footnotesize$\mathsf{b}$}
                (6);
        \end{tikzpicture}
        \caption{Extracted model for $\tup{\child}\tup{@_2\astep[b]2? =_\cmp \astep[b](q\land 3)?}$}
        \label{ex:extracted:model}
    \end{figure}
    Notice that in the extracted model nodes correspond to urfathers; in this respect, we have $\urf[7]=3$, $\urf[6]=2$, and for any other nominal $k$, $\urf[k]=k$.
\end{exa}

Up to now we focused on node expressions.  Handling satisfiability of path expressions is simple, given the following proposition. 

\begin{prop}
	Let $\alpha$ be a path expression of $\Hxpath$, then $\alpha$ is satisfiable if and only if $i{:}\tup{\alpha}\top$ is satisfiable, for $i$ a nominal not in $\alpha$. 
\end{prop}	

\begin{proof}
	The implication from right to left is obvious.  For the other direction, any model satisfying $\alpha$ can be turn into a model of $i{:}\tup{\alpha}\top$ by setting $\naming(i)$ to the starting point of the path satisfying $\alpha$. 
\end{proof}	
	
\section{Termination}\label{sec:termination}

In the previous section we show how tableaux yield a satisfiability calculus.
In this section, we prove that we can restrict these tableaux to be finite, without loosing completeness.
Such a result establishes that the satisfiability problem for $\Hxpath$ is decidable.

We begin by observing that the analyticity of the tableau ensures that all expressions introduced during expansion are constructed from subcomponents of the root, reducing the search space. 
The main technical challenge lies in managing the introduction of new nominals and new accessibility constraints through applications of the ($\Diamond$) and (child) rules. 
To resolve this, we adapt the termination argument for Hybrid Logic tableaux in~\cite{BolanderB07}. In particular, we impose a controlled strategy for applying the ($\Diamond$) and (child) rules to ensure that the resulting structure remains tree-like. 

\begin{defi}
  \label{def:atform}
  Let $\Theta$ be a branch of a tableau with root $i{:}\varphi$.
  For $j \in \noms(\Theta)$, define:
  \begin{align*}
    \at_{\nexpr}^{\Theta}(j) & = \setof{\psi}{{j{:}\psi} \in (\Theta \cap \qsub(i{:}\varphi))} \\
    \at_{\pexpr}^{\Theta}(j) & = \setof{\alpha}{\pm \<@_j\alpha \cmpr \beta \> \in (\Theta \cap \qsub(i{:}\varphi))}.
  \end{align*}
  where $\pm$ indicates that $\<@_j\alpha \cmpr \beta \>$ may or may not be preceded by a negation.
  In addition, define $\at^\Theta(j) = \at_{\nexpr}^{\Theta}(j) \cup \at_{\pexpr}^{\Theta}(j)$.
\end{defi}

Intuitively, the set $\at_{\nexpr}^{\Theta}(j)$ captures the subcomponents of the root of the tableau true at $j$ in the branch.
In turn, the set $\at_{\pexpr}^{\Theta}(j)$ captures the subpaths present in the data (in)equalities in the root of the tableau that start at $j$ in the branch.

\begin{exa} 
  \label{ex:at}
  Let $\Theta$ be the sole branch in the tableau in \Cref{ex:usetableaux}.
  Then
$$	      \at^{\Theta}(5) =       {\at_{\nexpr}^{\Theta}(5) \cup \at_{\pexpr}^{\Theta}(5)} =	      {\set{\tup{@_2\astep[b]2? =_\cmp \astep[b](q \land 3)?}} \cup \set{@_2\astep[b]2?,\astep[b](q\land 3)?}}.$$
  Notice how the set $\at^{\Theta}(5)$ gives us a hint of the rules which can be applied to the node expressions having $5$ as a prefix. 
  We retake this idea later on.
\end{exa}

The set $\qsub(i{:}\varphi)$ of quasi subcomponents of the label of the root of a tableau is infinite, since its considers all possible ways of prefixing it with nominals.
However, we can establish that in fact $\at^\Theta$ only considers those  quasicomponents that are made of either a subcomponent or a negated subcomponent, i.e., it is restricted to nominals appearing in a branch of the tableau. This is stated in the following result.

\begin{prop} \label{prop:at:finite}
  For every branch $\Theta$, the set $\at^\Theta(j)$ is finite.
\end{prop}
\begin{proof}
  Let $\Theta$ be a branch of a tableau with root $i{:} \varphi$.
  From \Cref{prop:analytic}, we get: 
  \begin{align*}
    {\at^{\Theta}_{\nexpr}}(j)
      & \subseteq  \setof{\psi, \lnot \psi}{\psi \in \sub(i{:}\varphi) } 
    \\
    \at^{\Theta}_{\pexpr}(j)
      & \subseteq \setof{\alpha}{{\pm}\tup{\alpha \cmpr \beta} \in \sub(i{:}\varphi)}.
  \end{align*}
  Since the set $\sub(i{:}\varphi)$ is finite, we immediately get that the set $\at^\Theta(j)$ is finite.
\end{proof}

\Cref{prop:at:finite} tells us that the set of expressions that are true at a given node must be finite. Still, there is no guarantee that every tableau must be finite. It could be the case that the tableau is infinitely branching, or that it has a branch of infinite size. We prove that neither is the case for our tableaux.

\begin{defi} \label{def:generated}
  Let $\Theta$ be a branch of a tableau, and $\{i, j\} \subseteq \noms(\Theta)$.
  We say that \emph{$j$ is generated by~$i$}, and write $i \prec_\Theta j$, iff there is $\pathh\in\rels$ such that $i{:}\tup{\pathh}j\in\Theta$ is an accessibility constraint. 
  Finally, we say that $i$ is an \emph{ancestor} of $j$, 
    and write $i \prec_\Theta^{+} j$,
  iff $(i,j)$ is in the transitive closure of $\prec_\Theta$.
\end{defi}

The following lemma tells us that every nominal in the branch of a tableau generates a finite number of new nominals. More precisely, this result tells us that applications of ($\Diamond$) and (child) result in a finite set of trees, each of which has a finite number of branches.

\begin{lem} \label{lemma:graph}
  Let $\Theta$ be a branch of a tableau.
  The graph $G_{\Theta} = (\noms(\Theta), \prec_\Theta)$ is a finite set of finitely branching trees.
\end{lem}
\begin{proof}
  That $G_{\Theta}$ is a finite set of trees follows from the following three facts.
  First, for every $\{i,j,k\} \subseteq \noms(\Theta)$, if $j \prec_\Theta i$ and $k \prec_\Theta i$, then, $j = k$, i.e., if a nominal is generated by another nominal, then it is generated by at most one other nominal.
  Second, the set of ancestors of a nominal $i \in \noms(\Theta)$ that is generated by another nominal has a least element.
  Third, for any $i \in \noms(\Theta)$, the set $\at^\Theta(i)$ is finite (see~\Cref{prop:at:finite}), this implies that there is at most $\card(\at^\Theta(i))$ new nominals generated from $i$.
\end{proof}

Now that we know that applications of ($\Diamond$) and (child) always result in a finite set of finitely branching trees, we can move on to show that every branch in such trees is of finite size. Such a result establishes that tableaux are finite. 

\begin{lem} \label{lemma:chain{:}infinite}
  A branch of a tableau is infinite iff there is an infinite chain $i \prec_\Theta j  \prec_\Theta \dots$.
\end{lem}

\begin{proof}
  The right-to-left direction is trivial.
  To prove the left-to-right direction, let  $\Theta$ be an infinite branch.
  From \Cref{prop:at:finite}, the branch $\Theta$ must contain infinitely many distinct nominals.
  This implies that $\noms(\Theta)$ in \Cref{lemma:graph} must be infinite.
  Since, from \Cref{lemma:graph}, we know that $G_\Theta$ is a finite set of finitely branching trees, it must be that some tree in $G_\Theta$ contains an infinite chain $i \prec_\Theta j \prec_\Theta \dots $.
\end{proof}

\Cref{def:maxsize} introduces the last piece needed to prove termination.

\begin{defi}\label{def:maxsize}
  Let $\Theta$ be a branch of a tableau, and let $i \in \noms(\Theta)$. Define
  \begin{align*}
    \maxnode(i) & = 
      \max\setof{ \size(\varphi)}{ \varphi \in \at^\Theta_{\nexpr}(i) } \\
    \maxpath(i) &=
      \begin{cases}
        \max\setof{\size(\alpha)}{\alpha \in \at^{\Theta}_{\pexpr}(i)} & \text{if } \at^{\Theta}_{\pexpr}(i) \neq \emptyset \\
        0 & \text{otherwise}\\
      \end{cases}\\
    \maxsize(i)
& = \max (\maxnode(i) , \maxpath(i)).
  \end{align*}
\end{defi}

Intuitively, $ \maxsize(i)$ indicates the size of the largest node expression true at $i$ in $\Theta$, or the length of the longest path expression starting at $i$.
The idea is to prove that such a measure of size decreases after the application of the tableaux rules. 

\begin{lem} \label{lemma:decreasing:length}
  Let $\Theta$ be a saturated branch. If $i \prec_\Theta j$, then $\maxsize(i) > \maxsize(j)$.
\end{lem}
\begin{proof}
  Let $i \prec_\Theta j$.
  We have three possible cases:
    (1) $\maxsize(j) = 0$,
    (2) $\maxsize(j) = \size(\psi)$ for some node expression $\psi$, and
    (3) $\maxsize(j) = \size(\alpha)$ for some path expression $\alpha$.
  The latter has two possible sub-cases:
    (a) $\tup{@_j\alpha \cmpr \beta} \in \Theta$ and $\lnot\tup{@_j\alpha \cmpr \beta} \notin \Theta$, and
    (b) $\lnot\tup{@_j\alpha \cmpr \beta} \in \Theta$ and $\tup{@_j\alpha \cmpr \beta} \notin \Theta$.
  The result is immediate for (1). 
  We prove the remaining cases below.
  
  \begin{description}
    \item[(Case 2)]
      Let $\maxsize(j) = \size(\psi)$ for some node expression $\psi$.
      We need to prove $\maxsize(i) > \size(\psi)$.
      %
        %
      It is clear that $j{:}\psi$ can only result from ($\Diamond$), ($\lnot\Diamond$), (ref), (sym), or (trans).
        If $j{:}\psi$ results from ($\Diamond$), using \Cref{lemma:graph}, we know $i{:}\tup{\child}\psi \in \Theta$.
        This implies, $\maxsize(i) \geq \size(\tup{\child} \psi) > \size (\psi) = \maxsize(j)$ as needed.
        In turn, if $j{:}\psi$ results from ($\lnot\Diamond$), we know $\psi = \lnot\psi'$.
        Using \Cref{lemma:graph} again, we know $\{i{:}\lnot\tup{\child}\psi', i{:}\tup{\pathh}j\} \subseteq \Theta$.
        This implies, $\maxsize(i) \geq \size(\lnot\tup{\child} \psi') > \size (\psi) = \maxsize(j)$ as needed.
        %
        Lastly, if $j{:}\psi$ results from (ref), (sym), or (trans), we have ${\size(\psi)=1}$. 
        At the same time, from $i \prec_{\Theta}j$, we have $\{i{:}\<\child\>\chi, \<\goto{i}\child\alpha \cmpr \beta\>\} \cap \Theta \neq \emptyset$.
        Immediately, $\maxsize(i) > 1 = \size(\psi)$ as needed.

  
    \item[(Case 3.a)]
      Let $\maxsize(j) = \size(\alpha)$, $\tup{@_j\alpha \cmpr \beta} \in \Theta$, and $\lnot\tup{@_j\alpha \cmpr \beta} \notin \Theta$.
      We need to prove $\maxsize(i) > \size(\alpha)$.
      %
      It is clear that $\tup{@_j\alpha \cmpr \beta}$ can only result from (child).
      Using \Cref{lemma:graph}, we know that $\tup{@_i\pathh\alpha \cmpr \beta} \in \Theta$.
      Then, $\maxsize(i) \geq \size(\tup{@_i\pathh\alpha \cmpr \beta}) > \size (\tup{\alpha \cmpr \beta}) = \maxsize(j)$ as needed.
    
    \item[(Case 3.b)]
      Let $\maxsize(j) = \size(\alpha)$, $\lnot\tup{@_j\alpha \cmpr \beta} \in \Theta$, and $\tup{@_j\alpha \cmpr \beta} \notin \Theta$.
      We need to prove $\maxsize(i) > \size(\alpha)$.
      %
      It is clear that $\lnot\tup{@_j\alpha \cmpr \beta}$ can only result from ($\lnot$child).
      Using \Cref{lemma:graph}, we know that $\lnot\tup{@_i\pathh\alpha \cmpr \beta} \in \Theta$.
      Then, $\maxsize(i) \geq \size({\lnot\tup{@_i\pathh\alpha \cmpr \beta}}) > \size (\lnot\tup{\alpha \cmpr \beta}) = \maxsize(j)$ as needed. \qedhere



  \end{description}
\end{proof}

\begin{lem} \label{prop:termination}	
  Let $\varphi$ be a node expression.
  Any tableau for $\varphi$ is finite.
\end{lem}
\begin{proof}
  The proof is by contradiction. Suppose there is an infinite tableau for $\varphi$. Then it must be that this tableau has an infinite branch $\Theta$.
  By \Cref{lemma:chain{:}infinite}, there exists an infinite chain
  $
   i \prec_\Theta j \prec_\Theta \dots  
  $
  At the same time, by \Cref{lemma:decreasing:length}, we have 
  $
    \maxsize(i) > \maxsize(j) > \dots  
  $
  This yields a contradiction; since for any nominal $i$, it its trivial that $\maxsize(i) \geq 0$.
\end{proof}

Combining the result of \Cref{prop:termination} with the completeness of our tableau-based satisfiability procedure, we obtain the main result of this section.

\begin{thm}\label{th:decidability}
  The satisfiability problem for $\Hxpath$ is decidable.
\end{thm}

In \Cref{sec:complexity} we refine the result in \Cref{th:decidability} and characterize the exact complexity of the satisfiability problem for $\Hxpath$. 
\section{Complexity of the Satisfiability Problem for $\Hxpath$}
\label{sec:complexity}

In this section, we prove that the satisfiability problem for $\Hxpath$ is \pspace-complete.
\pspace-hardness follows from \pspace-hardness of the Basic Modal Logic~\cite{blackburn2001modal,blackburn06}, a proper fragment of $\Hxpath$.
For completeness, we present an algorithm that applies the tableau rules in a controlled manner, ensuring that satisfiability checking can be performed in polynomial space.
Our algorithm explores the model induced by the tableau in a depth-first fashion.
This requires particular care, as this model is not necessarily tree-shaped and may contain cycles or shared substructures.
Furthermore, we make the necessary mechanisms explicit, making our algorithm suitable for implementation, a contribution by its own sake.

\begin{algorithm}[t]
    \caption{$\SATALG$}\label[algorithm]{alg:start}
    \begin{small}
        \Function{$\SATALG(\varphi : \text{node expression})$}
        {
            {\bf require:} true \\
            $\NOM \gets \text{a new nominal}$\;
            $\Phi\gets [\NOM{:}\varphi]$\;
            \Return{$\SATALG(\Phi,\NOM+1)$}\;
            {\bf ensure:} $\SATALG(\varphi)$ iff $\varphi$ is satisfiable \\
        }
    \end{small}
\end{algorithm}

\begin{algorithm}[t]
    \caption{$\SATALG$}\label[algorithm]{alg:sat}
    \begin{small}
        \Function{$\SATALG(\Phi : \text{list of node expressions}, \NOM : \text{nominal})$}
        {
            {\bf require:} the nominals in $\Phi$ are in $[0 \dots \NOM)$ \\


            \If{$\Phi = [~]$}{\KwRet{true}}

            \vspace*{2pt}

            $\BRANCH, \STACK \gets \Phi, [~]$\;
            \Repeat{$\Sigma = [~]$}
            {
                \If(\Comment*[f]{explore an alternative branch}){$\Sigma = [(\mathrm{b},\chi)] + \mathrm{L}$}
                {
                    $\BRANCH, \STACK \gets \BRANCH_{\text{b}} + [\chi], \mathrm{L}$;
                }

                \vspace*{2pt}

                \Expand(\Comment*[f]{close $\BRANCH$ under the application of rules of type 1})
                {
                    \If{there is a rule $\RULE$ of type 1 that is untreated in $\BRANCH$}
                    {
                        {\bf let} $\Psi$ be the consequent of $\RULE$\\
                        
                        $\BRANCH \gets \BRANCH + \Psi$;
                        \Comment*[f]{expand the branch}\\
                        \GoTo expand
                    }

                    \If{there is a rule $\RULE$ of type 2 that is untreated in $\BRANCH$}
                    {
                        {\bf let} $\{\psi, \chi\}$ be the consequent of $\RULE$\\
                        
                        $\BRANCH \gets \BRANCH + [\psi]$;
                        \Comment*[f]{expand the branch}\\
                        $\STACK \gets [(|\BRANCH|, \chi)] + \Sigma$;
                        \Comment*[f]{store the alternative and its location in the branch}\\
                        \GoTo expand
                    }
                }

                \vspace*{2pt}

                \If(\Comment*[f]{try an alternative branch}){there is a clash in $\BRANCH$ and $\Sigma \neq [~]$}
                    {{\bf continue}\;}
                \ElseIf(\Comment*[f]{no alternative branches to try}){there is a clash in $\BRANCH$}
                    {\Return{false}\;}

                \vspace*{2pt}

                \RecStep(\Comment*[f]{explore diamonds in a DFS manner})
                {
                    r $\gets$ true\;
                    \ForAll{rules $\RULE$ of type 3 applicable in $\BRANCH$}
                    {
                        {\bf let} $\psi$ be the premiss of $\RULE$ and $\psi \in \{i{:}\tup{\pathh}\chi, \tup{@_i\pathh\alpha \cmpr \beta}\}$\;

                        $\Phi$ $\gets$ $\Phi$ + $\BRANCH$ minus all diamonds in $\BRANCH$ + $[i{:}\tup{\pathh}\NOM]$\;

                        \If{$\psi = i{:}\tup{\pathh}\chi$}%
                            {$\Phi \gets \Phi + [\NOM{:}\chi]$}

                        \If{$\psi = \tup{@_i\pathh\alpha \cmpr \beta}$}%
                            {$\Phi \gets \Phi + [\tup{@_{\NOM}\alpha \cmpr \beta}]$}

                        {r $\gets$ r and $\SATALG(\Phi, \NOM+1)$}\;

                        $\Phi$ $\gets$ list of root literals and rooted boxes in $\Phi$\;
                    }
                }

                \vspace*{2pt}

                \lIf(\Comment*[f]{the branch is open and saturated}){r}{\Return{true}}
            }
            \KwRet{false}\;
            {\bf ensure:} $\SATALG([\NOM{:}\varphi],\NOM+1)$ iff $\varphi$ is satisfiable
        }
    \end{small}
\end{algorithm}

We begin by introducing some preliminary concepts. 
First, recall that without loss of generality, we assume that $\noms$ is the set of natural numbers. 
In what follows we use
$[~]$ for the empty list,
$[\mathrm{e}]$ for a list with a single element $\mathrm{e}$,
$\mathrm{L}_1 + \mathrm{L}_2$ for the concatenation of lists $\mathrm{L}_1$ and $\mathrm{L}_2$,
and
$|\mathrm{L}|$ for the length 
of a list $\mathrm{L}$.
Moreover, we use $\mathrm{L} = [\mathrm{e}]+\mathrm{L}'$ to indicate a non-empty list $\mathrm{L}$ with first element $\mathrm{e}$ and tail $\mathrm{L}'$, and $\mathrm{L}_{\mathrm{n}}$ to indicate $\mathrm{L}$'s initial segment of length $\mathrm{n}$.
Finally, we assume that parameter variables are pass-by-reference and assignments are copy constructors.

The algorithmic application of tableau rules starts by calling the function $\SATALG$ in \Cref{alg:start} with a node expression $\varphi$.
In this function, we set up the context for the application of tableau rules.
The actual application of rules is implemented in the overloaded $\SATALG$ in \Cref{alg:sat}. 

To keep the space complexity polynomial in \Cref{alg:sat}, we explore branches one at a time.
The information needed to explore branches is kept in local lists $\BRANCH$ and $\STACK$.
Intuitively, $\BRANCH$ stores the current branch of the tableau.
In turn, $\STACK$ behaves as a stack storing pairs $(\mathrm{b},\chi)$ where $\mathrm{b}$ is a number indicating the position to backtrack in $\BRANCH$ to explore an alternative branch, and $\chi$ is the alternative node expression to be explored.

Turning to branch expansions, we classify tableau rules into types 1, 2, and 3.
    The rules of type 1 are: the rules $({\lnot}{\lnot})$, $(\land)$, and $({\lnot}\Diamond)$ in \Cref{rules:basic};
    all the rules in \Cref{rules:nominals,rules:internalization};
    the rules (test), $(@)$, ($\lnot$child), $(\lnot@)$, $(\lnot{\cup})$, and (com$_1$), (com$_2$), and (dTrans) in \Cref{rules:path}; and 
    the rules
    \begin{align*}
        {\infer[(\mathrm{copy}_0)]
            {\phantom{|} k{:}\tup{\child}j}
            {\deduce[]
                {\phantom{|}i{:}j}
                {\phantom{|}k{:}\tup{\child}i}
            }
        }
        \qquad
        {\infer[(\mathrm{copy}_1)]
            {\phantom{|} \tup{@_j\alpha \cmpr \beta}}
            {\deduce[]
                {\phantom{|}i{:}j}
                {\phantom{|}\tup{@_i\alpha \cmpr \beta}}
            }
        }
        \qquad
        {\infer[(\mathrm{copy}_2)]
            {\phantom{|} \lnot\tup{@_j\alpha \cmpr \beta}}
            {\deduce[]
                {\phantom{|}i{:}j}
                {\phantom{|}\lnot\tup{@_i\alpha \cmpr \beta}}
            }
        }
    \end{align*}
    These rules have side conditions $j < i$ and $j$ a root nominal.
    In brief, these rules may be understood in analogy with (copy) in \Cref{rules:nominals}.
    These rules are admissible and do not affect termination.
    They are introduced here only to cope with space complexity.
    %
    The rules of type 2 are: the rule $(\lnot{\land})$ in \Cref{rules:basic}, and the rules ($\lnot$test), and $({\cup})$ in \Cref{rules:path}.
    %
    Finally, the rules of type 3 are: the rule $(\Diamond)$ in \Cref{rules:basic}, and
        the rule (child) in \Cref{rules:path}.
The rationale behind the classification of tableau rules is that rules of type 1 extend the branch linearly, rules of type 2 split the branch into two, and rules of type 3 handle diamonds.

We are now in a position to explain how~\Cref{alg:sat} applies tableau rules to the current branch, i.e., to the list $\BRANCH$.
It is clear that the application of a rule should expand $\BRANCH$ by adding node expressions at its end.
This expansion is actually dealt with in two blocks of code:
\begin{description}
    \item[(expand)]
        Lines 9 to 18.
        This block of code applies rules of type 1 or of type 2 until there are no untreated rules.
        The case of a rule $\RULE$ of type 1 is straightforward.
        Let $\Psi$ be the set of node expressions in the consequent of $\RULE$.
        We say that $\RULE$ is untreated iff $\Psi \nsubseteq \BRANCH$.
        If $\RULE$ is untreated, we add all the node expressions in $\Psi$ at the end of $\BRANCH$. 
        In turn, let $\RULE$ be a rule of type 2 and let $\Psi = \{\psi,\chi\}$ be the node expressions in the consequent of $\RULE$. 
        We say that $\RULE$ is untreated iff $\Psi \cap \BRANCH = \emptyset$.
        If $\RULE$ is untreated, we add $\psi$ at the end of $\BRANCH$, and store in $\STACK$ the alternative $\chi$ together with the point in $\BRANCH$ at which $\psi$ is introduced.

    \item[(explore)]
        Lines 23 to 33.
        This block deals with the exploration of ``diamonds'', i.e., node expressions of the form $i{:}\tup{\pathh}\psi$ that are not accesibility constraints, or node expressions of the form $\tup{@_i\pathh\alpha \cmpr \beta}$. 
        To do so, we take some ideas from \cite{Ladner77,Marx06} of exploring diamonds in a depth first search (DFS) manner and adapt them to our setting.
        
        First, in our case, the exploration of diamonds corresponds to the application of rules of type 3.
        A rule $\RULE$ of type 3 is applicable iff its premiss belongs to $\BRANCH$.
        If $\RULE$ is applicable, its application involves: creating a new nominal $\NOM$, creating the accesibility constraint $i{:}\tup{\child}\NOM$, and adding the node expression in the consequent of $\RULE$ to the branch.

        Second, exploring diamonds in a DFS manner involves recursive calls to $\SATALG$.
        These recursive calls take into account all alternative explorations of diamonds and put a temporary pause on them. Notice that only one diamond must be treated at each step, as they can be individually satisfied and only can find contradictions with box expressions. Once the satisfiability of one particular diamond is checked, the algorithm backtracks and all diamonds are restored on $\BRANCH$. 
        This is implemented by removing all alternative diamonds from the branch in line 27 of \Cref{alg:sat}, before the recursive call.

        Lastly, before backtracking in the DFS strategy we may need to gather \emph{global} information to be passed along to alternative calls to $\SATALG$.
        This global information 
        is mediated by root nominals in a tableau branch and corresponds to \emph{root literals} and \emph{rooted boxes}.
        By a literal, we mean a node expression of the form $i{:}j$, $i{:}\lnot j$, $i{:}p$, $i{:}\lnot p$, $i{:}\tup{\child}j$, $\tup{@_i \cmpr @_j}$ where $\{i,j\} \subseteq \noms$ and $p \in \props$.
        A root literal is one whose nominals are root nominals.
        Root literals contain the relevant information stored in urfathers of root nominals, thus they need to be passed to the next recursive call.
        A rooted box is a node expression of the form $i{:}\lnot\tup{\child}\psi$ or $\lnot\tup{@_i\alpha \cmpr @_j\beta}$ where $i$ and $j$ are root nominals.
        Global information to the branch is recorded in line 33 of \Cref{alg:sat}.
\end{description}

Correctness of \Cref{alg:sat} follows from the correctness of tableau rules, while termination is ensured since each loop and recursive call is bounded. The next result shows the algorithm uses only polynomial space.

\begin{thm}
    \label{th:satpspace}
    The satisfiability problem for $\Hxpath$ is \pspace-complete.
\end{thm}
\begin{proof}
    Hardness follows from the fact that the Basic Modal Logic~$\K$ is \pspace-hard~\cite{blackburn2001modal,blackburn06}, and $\Hxpath$ subsumes it. 
    Completeness follows from the fact that \Cref{alg:sat} uses at most polynomial space with respect to the size of the input node expression, as analyzed in what follows. 

    The local variables $\BRANCH$ and $\STACK$ occupy at most a polynomial amount of memory (bounded by the number of untreated rules of type 1 and 2).
    The number of recursive calls, in a branch, is also polynomial (bounded by the number of nested diamonds, including path expressions of the form $\child$). 
    The only information that is global to the branch is that on the list $\Phi$.
    But this list is also of polynomial size, as the only persistent information that is passed along between different recursive calls refers only to root literals, and this information is polynomial.

    When put together, these facts tell us that \Cref{alg:sat} uses at most a polynomial amount of memory with respect to the size of the input node expression.
\end{proof}

We close this section with a run of the function $\SATALG$.
This run illustrates the inner workings of the algorithm it implements.
Moreover, it allows us to recognize how it uses only polynomial space.

\begin{exa}
Let us suppose that we start $\SATALG$ in~\Cref{alg:start} with the node expression
\[
    \varphi = \tup{\child(1\land p)? =_\cmp \astep[b]} \land \tup{\astep[c]@_1(\neg p)? =_\cmp \astep[b]}  \land \neg\tup{\astep[b]\neq_\cmp @_1}.
\]   
The call to $\SATALG(\varphi)$ invokes $\SATALG(\Phi,\NOM)$ with $\Phi=[2{:}\varphi]$ and  $\NOM=3$.
We proceed to list the different steps in~\Cref{alg:sat} on this input.
Sometimes we group several steps together and omit some node expressions to make the presentation more succinct. 

\begin{description}
  \item[Initialization] 
    We set $\BRANCH=\Phi$ and $\STACK=[~]$.
    Notice that the sole root nominal in $\varphi$ is $1$.

  \item[Step 1] 
    We reach the `expand' block and apply rules of type 1 until no such rule is left untreated.
    The list $\BRANCH$ is updated during this process (the rules applied to obtain the node expressions are indicated on the right):
    \[
      \begin{array}{r@{\,}@{}lr}
        \BRANCH =
          & [
              2{:}\tup{\child(1\land p)? =_\cmp \astep[b]},
              2{:}\tup{\astep[c]@_1(\neg p)? =_\cmp \astep[b]},
              2{:}\neg\tup{\astep[b]\neq_\cmp @_1},
          ] + \dots +
              & (\land) \\
          & [
              \tup{@_2\child(1\land p)? =_\cmp @_2\astep[b]},
              \tup{@_2\astep[c]@_1(\neg p)? =_\cmp @_2\astep[b]},
              \neg\tup{@_2\astep[b]\neq_\cmp @_2@_1}
            ]
              & (\text{int}_1), (\text{int}_2)
      \end{array}
    \]

  \item[Step 2]
    We reach the `explore' block and choose to apply the rule of type 3 with premiss $\tup{@_2\child(1\land p)? =_\cmp @_2\astep[b]}$.
    Before the recursive call we get
    \[
      \Phi =
        [
          \neg\tup{@_2\astep[b]\neq_\cmp @_2@_1}, 2{:}\tup{\astep[a]}3, \tup{@_3(1\land p)? =_\cmp @_2\astep[b]}
        ]
    \]

  \item[Step 3] 
    We call $\SATALG(\Phi,\NOM)$ recursively for a first time, with $\NOM=4$. 
    In the `expand' block of the first recursive call, we apply all rules of type 1 and get
    \[
      \begin{array}{r@{\,}@{}lr}
        \BRANCH = 
          &
          [
            \neg\tup{@_2\astep[b]\neq_\cmp @_2@_1},
            2{:}\tup{\astep[a]}3,
            \tup{@_3(1\land p)? =_\cmp @_2\astep[b]},
          ] + \dots + \\
          &
          [
            3{:}1,3{:}p,1{:}3,1{:}p, 2{:}\tup{\astep[a]}1,
            \tup{@_3 =_\cmp @_2\astep[b]},
            \tup{@_1 =_\cmp @_2\astep[b]},
            \tup{@_2\astep[b]=_\cmp@_1}
          ]
      \end{array}
    \]
    Notice that $1{:}p$ is obtained from (copy), $2{:}\tup{\child}1$ from (copy$_0$) and $\tup{@_1=_\cmp @_2\astep[b]}$ from (copy$_1$). 
    This way, the nominal $3$ is replaced by the root nominal $1$.
    Finally, the last node expression in $\BRANCH$ is obtained by applying the rule (com$_1$).

  \item[Step 4]
    In the `explore' block of the first recursive call we choose to apply the rule of type 3 with premiss $\tup{@_2\astep[b]=_\cmp@_1}$.
    Before the recursive call we get
    \[
      \Phi =
        [
          \neg\tup{@_2\astep[b]\neq_\cmp @_2@_1}, 1{:}p, 2{:}\tup{\astep[a]}1, 2{:}\tup{\astep[b]}4, \tup{@_4 =_{\cmp} @_1}
        ]
    \]

  \item[Step 5]
    We call $\SATALG(\Phi,\NOM)$ recursively for a second time (with $\NOM=5$).
    In the `expand' block of the second recursive call, we apply all rules of type 1 and get
    \[
      \begin{array}{r@{\,}@{}lr}
        \BRANCH = 
            &
            [
              \neg\tup{@_2\astep[b]\neq_\cmp @_2@_1},
              2{:}\tup{\astep[b]}4,
              \tup{@_4 =_{\cmp} @_2@_1}
            ] + \dots + \\
            &
            [
              \neg \tup{@_4 \neq_{\cmp} @_2@_1}
            ] + \dots +
              & \text{($\lnot$child)} \\
            &
            [
              \neg \tup{@_4 \neq_{\cmp} @_1}
            ]
              & \text{($\neg@$)}
      \end{array}
    \]

  \item[Step 6] 
    Since no rules of type 2 are triggered, no clash is found, and there are some untreated node expressions, the algorithm proceeds to the `explore' phase again. A new rule of type 3 is triggered, this time with premise $\psi= \tup{@_2\astep[c]@_1(\neg p)? =_\cmp @_2\astep[b]}$. Before the new recursive call, we get
    \[
      \Phi =
        [
          \neg\tup{@_2\neq_{\cmp}@_1}, 1{:}p
        ] 
        + \dots + 
        [
          2{:}\tup{\astep[c]}5, \tup{@_5@_1(\neg p)? =_\cmp @_2\astep[b]} 
        ]
    \]

  \item[Step 7] 
    The recursive call invokes $\SATALG(\Phi,\NOM)$, with $\NOM=6$. In the `expand' block, we get 
    \[
      \begin{array}{r@{\,}@{}lr}
        \BRANCH = 
            &
            [
              \neg\tup{@_2\neq_{\cmp}@_1}, 1{:}p, 2{:}\tup{\astep[c]}5, \tup{@_5@_1(\neg p)? =_\cmp @_2\astep[b]}
            ] + \dots + \\
            &
            [
              \tup{@_1(\neg p)? =_{\cmp} @_2\astep[b]}, \tup{@_1 =_{\cmp}@_2\astep[b]}, 1{:}\neg p
            ] 
              & (@), (?) 
      \end{array}
    \]
    \item[Step 8] 
      After executing the Step 7, the algorithm finds a clash in line 21, since both $1{:}p$ and $1{:}\neg p$ belong to the sole tableau branch. Therefore, \Cref{alg:sat} returns that $\varphi$ is unsatisfiable.
\end{description}

Notice how in each step the algorithm keeps track of the list of accesibility constraints created in the exploration of diamonds in a DFS manner.
The length of these lists is bounded by the lengths of the path expressions traversed in each diamond.
Moreover, we only need to keep track of a polynomially bounded number of node expressions at each point in these paths.
This example illustrates the polynomial space bound in \Cref{th:satpspace}.
\end{exa}
\section{Forest and Tree Models}
\label{sec:extensionplus}

In this section, we extend our tableau calculus to ensure that extracted models are \emph{data trees}.
Besides their theoretical interest, date trees are relevant as they are the underlying structure of XML documents in practice.
In this extension of the calculus, we take advantage of the expressive power of nominals and the possibility to ``jump'' to evaluation points to characterize \emph{forests}, i.e., disjoint unions of trees, and trees as a particular case.
We introduce rules and clash conditions to determine whether nodes have at most one predecessor, and to avoid loops via accessibility relations.  
We then extend our soundness and completeness results to deal with the
class of forests.
Finally, we show how to  get completeness with respect to the class of trees by
checking a single constraint on open branches.

To work with forests and trees, we extend our tableaux rules with a new type of accessibility constraint called {\em transitivity constraint}.
Transitivity constraints are expressions of the form $i{:}\tup{+} j$.
They are introduced into the calculus to represent reachability navigating accessibility relations.
We introduce transitivity constraints and the new rules of the calculus in~\Cref{rules:reachability}.  
In this figure, the rule (+intro) ensures that the relation described by ${+}$ contains the relation described by ${\pathh}$, whereas the rule (trans) indicates that the relation described by ${+}$ is a transitive relation.
Finally, the rule (+copy) takes care of dealing with urfathers, copying transitivity constraints where they need to be copied.
 
\begin{figure}[t]
    \begin{center}
        \small 
        {\def\arraystretch{1.2}
        \begin{tabular}{c}
            \toprule
            {\sc Reachability Rules}
            \tabularnewline
            \midrule
            {
            \begin{tabular}{ccc}
                {\infer[(+\mathrm{intro})]
                { i{:}\tup{+} j}
                { i{:}\tup{\pathh} j}
                }
                &
                {\infer[(\mathrm{trans})]
                    {i{:} \tup{+} k}
                    {\deduce[]
                    { j{:}\tup{+} k}
                    {i{:}\tup{+} j} 
                    }
                }
                &
                {\infer[(+\mathrm{copy})]
                    {i{:}\tup{+} k}
                    {\deduce[]
                    {\phantom{|} j{:}k} 
                    {\phantom{|} i{:}\tup{\pathh} j}}
                }
            \end{tabular}
            }
            \tabularnewline
            \bottomrule                        
        \end{tabular}}
    \end{center}
    \caption{Tableau Rules for Forests and Trees.}
    \label{rules:reachability}
\end{figure}

\begin{rem}
     It is important to point out that even though transitivity constraints can be mistaken as characterizing the transitive closure of $\bigcup_{\pathh \in \rels} R_{\pathh}$, this is not the case. The relation associated to $\tup{+}$ in the extracted model may not be the smallest transitive relation containing $\bigcup_{\pathh \in \rels} R_{\pathh}$.
    In any case, transitivity constraints suffice for handling forests (and trees as a particular case).
    
    Furthermore, we can always assume that the symbol $+$ is not part of the set $\rels$, and then does not occur in the node expression at the root of the tableau. Thus, the obtained tableau is still internalized, but over the extended signature $\rels'=\rels\cup\{+\}$.
\end{rem}


In addition to the rules in~\Cref{rules:reachability}, we introduce new clash conditions for forests and trees.

\begin{defi}\label{def:clashf}
    A branch $\Theta$ is said to contain a \emph{clash} if it satisfies any of the conditions listed in \Cref{def:clash}, or if any of the following situations arise:
    \begin{enumerate}
        \setcounter{enumi}{2}
        \item $i{:}\tup{+}i \in \Theta$, 
        \item $\{j{:}\tup{+} i, k{:}\tup{+} i\} \subseteq \Theta$, and $\{j{:}k,j{:}\tup{+}k\}\cap\Theta = \emptyset$. 
    \end{enumerate}
\end{defi}

Notice how condition (3) in \Cref{def:clashf} is an indication that no node can be reached by itself; whereas condition (4) in \Cref{def:clashf} tells us that a node cannot have more than one predecessor. 
More precisely, condition (4) tells us that if a nominal is reachable from two others, then, they either indicate the same node, or one must be reachable from the other.

The following example demonstrates how the introduction of new rules and clash conditions restricts the range of models that satisfy a given node expression.

\begin{exa}
    \Cref{fig:btableau} depicts a tableau for $\tup{@_0 \pathh {0?} =_{\cmp} p?}$.
    In brief, $\tup{@_0\pathh 0? =_{\cmp} p?}$ forces a self loop at the node named by $0$.
    Applying only the rules in \Cref{rules:basic,rules:nominals,rules:internalization,rules:path} from \Cref{sec:tableaux} yields the first 11 nodes of the tableau.
    Notice that if we restrict our attention only to those 11 nodes (i.e., if we truncate the tableau at the node 11), we obtain a branch that is open and saturated according to \Cref{def:saturated} (modulo the application of reflexivity and symmetry rules, which are omitted).
    This indicates that $\tup{@_0\pathh 0? =_{\cmp} p?}$ is satisfiable in the class of all models.
    The nodes 12 and 13 show the effect of the new rules, and the way in which they allow us to talk about reachability.
    Notice also that node 13 introduces a clash condition as per \Cref{def:clashf}; i.e., a node is reachable from itself.
    This clash condition tells that $\tup{@_0 \pathh {0?} =_{\cmp} p?}$ is unsatisfiable in the class of tree models; and thus also unsatisfiable in the class of forests.
    \begin{figure}[!h]
        \begin{center}
        \begin{prooftree}{}%
        [
            1{:}\tup{@_0 \pathh {0?} =_{\cmp} p?},
            just={root}
            [
                \tup{@_1@_0 \pathh {0?} =_{\cmp} @_1p?},
                just={(int$_1$):!u}
                [
                    \tup{@_0 \pathh {0?} =_{\cmp} @_1p?},
                    just={($@$):!u}
                    [
                        0{:}\tup{\pathh}2,
                        just={(child):!u}
                        [
                            \tup{@_2{0?} =_{\cmp} @_1p?},
                            just={(child):!uu}
                            [
                                2{:}0,
                                just={(test):!u}
                                [
                                    \tup{@_2 =_{\cmp} @_1p?},
                                    just={(test):!uu}
                                    [
                                        \tup{@_1p? =_{\cmp} @_2},
                                        just={(com$_1$):!u}
                                        [
                                            1{:}p,
                                            just={(test):!u}
                                            [
                                                \tup{@_1 =_{\cmp} @_2},
                                                just={(test):!uu}
                                                [
                                                    \tup{@_2 =_{\cmp} @_1},
                                                    just={(com$_1$):!u}
                                                    [%
                                                                                    0{:}\tup{+}2,
                                                                                    just={($+$intro):!uuuuuuuu}
                                                                                    [
                                                                                        0{:}\tup{+}0,
                                                                                        just={($+$copy):!u,!uuuuuuu},
                                                                                        close={:!u}
                                                                                    ]
                                                    ]
                                                ]
                                            ]
                                        ]
                                    ]
                                ]
                            ]
                        ]
                    ]
                ]
            ]
        ]
        \end{prooftree}
        \end{center}
       \caption{A closed tableau for $\tup{@_0 \pathh {0?} =_{\cmp} p?}$}\label{fig:btableau}
    \end{figure}
\end{exa}

We are now in a position to prove that the new rules and clash conditions indeed force forest-like structures.

\begin{prop}
\label{prop:forest}
Let $\Theta$ be an open saturated branch in the extended calculus, and let $\model^{\Theta}$ be its extracted model.  
Then, $\model^{\Theta}$ is a forest.	
\end{prop}

\begin{proof}
    The proof is concluded if $F = (\bigcup_{\pathh\in\rels} R_{\pathh}^{\Theta})^{+}$ is a forest.
    To this end, we show that:
    \begin{enumerate}[(a)]
    \item it is not the case that $(n,n) \in \mathbin{F}$, and
    \item if $(n,m)\in \mathbin{F}$ and $(n',m) \in \mathbin{F}$, then $n = n'$ or $(n,n') \in \mathbin{F}$.
    \end{enumerate} 

    The proof of (a) is by contradiction.
    Suppose that $(n,n)\in \mathbin{F}$.
    Then, exist $\astep[a]_1,\ldots,\astep[a]_{\ell}$ such that
        $(n,i)\in R_{\astep[a]_1}, (i,j)\in R_{\astep[a]_2} \ldots (k,n)\in R_{\astep[a]_{\ell}}$.
    From the definition of $\cra$, we have that ${\{n{:}\tup{\pathh_1} i, i{:}\tup{\pathh_2}j,\ldots,k{:}\tup{\pathh_{\ell}} n\}\subseteq\Theta}$.
    Since $\Theta$ is closed under (+intro), we also have that $\{n{:}\tup{+} i, i{:}\tup{+}j,\ldots,k{:}\tup{+} n\}\subseteq\Theta$.
    In turn, closure under (trans), gives us ${n{:}\tup{+} n}\in \Theta$.
    This yields a clash using condition (3) from~\Cref{def:clashf}, contradicting the assumption that $\Theta$ is open.

    The proof of (b) is also by contradiction.
    Suppose $(n,m)\in \mathbin{F}$, $(n',m) \in \mathbin{F}$, $n \neq n'$, and $(n,n')\notin\mathrel{F}$.
    From $(n,m)\in \mathbin{F}$ and $(n',m)\in \mathbin{F}$, we know that $\{n{:}\tup{+}m,n'{:}\tup{+}m\} \subseteq \Theta$.
    In addition, from $(n,n')\notin \mathrel{F}$, we know $n{:}\tup{+}n' \notin \Theta$.
    Using the definition of urfather, we know that $j{:}k \notin \Theta$.
    When put together, these facts yield a clash using condition (4) from~\Cref{def:clashf}, contradicting the assumption that $\Theta$ is open.
\end{proof}

To further ensure that the resulting model is a tree we only need 
to verify that every point in the extracted model is reachable from 
the root. 

\begin{defi}
    \label{def:connected}
    Let $\Theta$ be a branch of a tableau for $i{:}\varphi$ in the extended calculus. We say that $\Theta$
    is {\em connected} iff for all $j\in\Theta$, if $i{:}j \notin \Theta$, then, $i{:}\tup{+}j \in \Theta$.
\end{defi}

\begin{prop}
    \label{prop:connectedtree}
    Let $\Theta$ be an open, saturated, and connected branch. Then, $\model^{\Theta}$ is a tree. 
\end{prop}

\begin{proof}
    The proof is concluded if $T = (\bigcup_{\pathh\in\rels} R_{\pathh}^{\Theta})^{+}$ is a tree.
    We already know that $T$ is a forest.
    Thus, we only need to prove that for all $j\in\cdom$, $(i,j)\in \mathbin{T}$.
    This is straightforward since $\Theta$ is connected.
\end{proof}

We conclude this section by showing that the satisfiability problem for forests and data trees remains decidable and stays within \pspace.

\begin{thm}
    \label{th:terminationtrees}
   The satisfiability problem for forests and data trees in $\Hxpath$ is decidable and \pspace-complete.
\end{thm}

\begin{proof}
    Notice that for forests and trees, we only add three new rules which are applied over
    a finite set of node expressions (by~\Cref{prop:termination}). Since we apply rules only once, this yields again a finite tableau. Moreover, the new rules are clearly of type 1, thus applying them in~\Cref{alg:sat} yields similar bounds as in~\Cref{th:satpspace}. Checking the clash conditions can also be done using polynomial space. Therefore, the extended algorithm runs in polynomial space.
\end{proof}

\section{Pure Axioms and Node Creating Rules}
\label{sec:pure-ext}

A common advantage of proof systems for Hybrid Logics is their ability to establish a very general completeness result, in which it is shown that certain extensions of the base proof system are automatically complete for the corresponding classes of models.  This is in sharp contrast with many proof systems for other Modal Logics, in which simple extensions of a complete proof system might be incomplete.

This property of Hybrid Logics is particularly useful when designing proof systems for specific classes of models. For instance, certain applications may require structural constraints ---such as transitivity or irreflexivity--- on accessibility relations.
This is specially relevant for XPath, as the different axes are usually associated with semantics (e.g., price, sub-category, sub-menu, max-level, etc.) which might need to obey structural restrictions. 
Extensions using pure axioms and node-creating rules address these requirements naturally, while still yielding automatic completeness results.

Extensions for $\Hxpath$ that preserve completeness have been investigated in~\cite{ArecesF21} for Hilbert-style axiom systems. In this section, we adopt a similar approach to develop complete tableau systems for extensions of $\Hxpath$. To this end, we follow the methodology introduced in~\cite{Blackburn00,Blackburn2002BeyondPA,BtC06} for Hybrid Logics. We will consider two kinds of extension: pure axioms, and node creating rules.  These extensions ensure the soundness and completeness of the resulting calculi, but do not generally preserve decidability or complexity bounds. Such results depend on carefully controlling the flow of information within the tableaux and must be established on a case-by-case basis. While some structural properties can be managed using the mechanisms introduced earlier, others may fall outside the scope of such approach.

\subsection{Pure Axioms} 
One of the most standard extensions of hybridized proof systems is via pure axioms. This extension enables us to characterize several interesting classes of models, including some that are not definable by so-called ``orthodox'' node expressions (i.e., expressions containing only propositional symbols). 

\begin{defi}\label{def:pure}
    A node expression is \emph{pure} iff it does not contain propositional symbols.
\end{defi}

\begin{exa}\label{ex:pure}
 Below we list some pure node expressions defining interesting frame conditions:
 \[
 \begin{array}{l@{\quad\quad}l}
    \text{Irreflexivity}  &  i{:}\neg \tup{\child} i \\
    \text{Transitivity} &  i{:}(\tup{\child}\tup{\child}j \ra \tup{\child}j) \\
    \text{Trichotomy} & i{:}\tup{\child} j \lor i{:}j \lor j{:}\tup{\child}i\\
    \text{Data Uniqueness} & \tup{@_i =_\cmp @_j} \to i{:} j .
 \end{array}
 \]
 It is interesting to notice that transitivity is definable without the use of nominals,  but trichotomy and irreflexivity are not (see \cite{Blackburn00}). The latter is also true for data uniqueness.
\end{exa}

It has been shown in~\cite{ArecesF21} that pure expressions correspond to first-order properties when taken as axioms. This makes them suitable to characterize interesting classes of models. Below we formally introduce such a notion.

\begin{defi}
    A \emph{(data) frame} is a tuple $\aframe = \tup{N, \set{R_\pathh}_{\pathh \in \rels}, \set{\approx_{\cmp}}_{\cmp\in\cmps}}$, where each component is as in~\Cref{def:models}.
    We write $\aframe\vDash\varphi$ iff for all $\model$ based on $\aframe$, and $n \in N$, we have $\model,n \vDash \varphi$.
    We say that a node expression~$\varphi$ \emph{defines a class of frames $\fframe$} iff $\aframe\vDash\varphi$ is equivalent to  $\aframe\in\fframe$. 
\end{defi}

Our tableaux receive as input node expressions that are prefixed by some $i{:}$, so in order to handle pure axioms we need to make sure they are also prefixed. Following~\cite{Blackburn00,Blackburn2002BeyondPA}, in the following lemma we show that this is not a real restriction.

\begin{lem}\label{lemma:defines}
 If a pure node expression $\varphi$ defines a class of frames $\fframe$ and $i$ a nominal not in $\varphi$, then, $i{:}\varphi$ also defines~$\fframe$.
\end{lem}
\begin{proof}
    The result follows immediately from the requisite that $i$ does not appear in $\varphi$, as in this case $i$ does not affect the meaning of $\varphi$.
\end{proof}

We are now in position to state a general completeness result. The idea is to incorporate (prefixed) pure expressions to check satisfiability of a node expression in a particular class of models. In other words, if we look for a model of a given node expression, pure axioms will guarantee that this model will belong to the corresponding class of models. Given a set of pure expressions $\axioms$ and $\psi\in\axioms$, we need to instantiate nominals appearing in $\psi$, with nominals appearing in the branch of the tableaux in which~$\psi$ is being incorporated. For instance, for a branch $\Theta$ with $k{:}\varphi\in\Theta$, we incorporate irreflexivity as formulated in~\Cref{ex:pure} with the instance $k{:}\neg\tup{\child}k$. This kind of remark applies also to data comparisons. 

\begin{thm}[Strong Completeness for Pure Axioms]\label{th:strong}
    \label{th:pure-completeness}
    Let $\axioms$ be a set of pure node expressions prefixed by some $i{:}$.
    Extend the tableau calculus from~\Cref{sec:tableaux} with the rule (Pure) which adds, at any stage, any $\psi\in\axioms$ instantiated with nominals already appearing in the branch. The obtained tableau is complete with respect to the class of models based on frames defined by $\bigwedge_{\psi\in\axioms} \psi$.
\end{thm}

\begin{proof}
 The argument for this proof is the same as the one in~\cite[Th.~7.2]{Blackburn00} and \cite[Prop.~3.2]{Blackburn2002BeyondPA} for tableau calculi for hybrid logics, and~\cite[Th.~3.23]{ArecesF21} for axiomatizations of hybrid XPath.  Intuitively, the rule (Pure) ensures that pure axioms in $\axioms$ are fully instantiated in a suitable way, in an open and saturated branch $\Theta$. As a consequence, the extracted model $\model^\Theta$ globally satisfies all relevant instances in $\axioms$. This is sufficient to ensure that the frame of $\model^\Theta$ satisfies the structural properties associated with $\axioms$. In 
 other words, the extracted model $\model^\Theta$, satisfying the root formula in the branch, will belong to the required model class. 
\end{proof}

By choosing a suitable set of pure axioms it is possible to characterize interesting XPath features. 
We introduce some of them in the following example.

\begin{exa}\label{ex:extensions}
As our first example of the use of pure actions, we begin by considering the  \emph{inverse} relation, a distinctive feature in XPath fragments which allows for backwards navigation. To this end, we extend the set of relation symbols in the signature to $\rels'=\rel\cup\set{\child^- \mid \child \in \rels}$. 
Then, we define
\[
     \axioms[Inverse] = \{i{:}(j\to[\child]\tup{\child^-}j),i{:}(j\to[\child^-]\tup{\child}j) \mid \child \in \rels\}.
\]
By incorporating $\axioms[Inverse]$, our tableaux yields a complete calculus for $\Hxpath$ with inverse relations (over the new signature). More precisely, pure axioms in $\axioms[Inverse]$ ensure that, in any model considered, $R_\textsf{a} = (R_{\textsf{a}^-})^-$. Hence, $\textsf{a}^-$ can be used to refer to the inverse of 
$\textsf{a}$. 

For a second example, fix now $\rels''=\rels'\cup\set{\sib}$, and define
\[
     \axioms[Sibling] = \axioms[Inverse] \cup \{i{:}(\tup{\sib}j\to \tup{\child^-}\tup{\child}j), i{:}(\neg j \land \tup{\child^-}\tup{\child}j \to \tup{\sib}j), i{:}(j{:}\neg\tup{\sib}j) \mid \child \in \rels\}.
\]
By incorporating $\axioms[Sibling]$ we get, additionally, completeness for $\Hxpath$ with $R_{\sib}$ being the \emph{sibling} relation. The first node expression states that a sibling is one node found going backwards and then forwards in the model. The second node expression states the converse of the first expression, in case $j$ is not the current node. Finally, the last expression expresses that a node is not its own sibling. 

As a final example, define 
\[
    \axioms[\sib\text{-}Irreflexivity] = \axioms[Sibling] \cup \{ i{:}(\tup{\sib} j \to (\tup{@_i \neq_\cmp @_j})) \}.  
\]
$\axioms[\sib\text{-}Irreflexivity]$ gives us completeness with respect to the class of models in which siblings have different data values for the comparison criteria $\cmp$. This could be useful, for example, if when the siblings in the data tree represent categories with different prize ranges (e.g., economy, business, first-class).
\end{exa}

We stress that the Strong Completeness Theorem (\Cref{th:strong}) guarantees that the resulting calculi are complete for the intended model classes.  All the extensions discussed in~\Cref{ex:extensions} correspond to common features found in various versions of XPath. It is therefore desirable to have tableau calculi that are not only sound but also complete for these extensions. As it turns out, in all these cases, completeness is automatically preserved due to the specific form of the axioms being added. This being said, adding such axioms does not imply complexity bounds are preserved.  
For instance, the satisfiability problem for Hybrid Logic $\mathcal{H}(:)$ extended with the inverse operator is already known to be \textsc{ExpTime}-complete~\cite{Areces2000c}. Since $\mathcal{H}(:)$ with inverses is a proper fragment of $\Hxpath$ extended with inverses, this immediately yields an \textsc{ExpTime} lower bound for the satisfiability problem of $\Hxpath$ with inverses.

\subsection{Node Creating Rules} We now turn our attention to introducing so-called \emph{node creation rules} into our tableaux. To precisely characterize the expressivity provided by these new rules, we need to extend the basic language $\Hxpath$ to a strong version with quantification over nominals, as is done for Hybrid Logic~\cite{BlackburnS95}. It is important to clarify that the strong hybrid language for XPath introduced in~\Cref{def:strong} is used solely to characterize the expressive power of the node-creating rules defined below. It is also worth noting that the satisfiability problem for the strong hybrid language for XPath is undecidable~\cite{arec:hybr05b}. Therefore, node-creating rules must be used with caution when aiming for a tableau calculus with good computational properties. 

\begin{defi}[Strong Hybrid Language for XPath]\label{def:strong}
    \label{def:stronghl}
    The \emph{strong hybrid language} for XPath is obtained by allowing expressions of the form $\exists x.\varphi$ (where $x\in\noms$, and $\varphi$ is a node expression) as node expressions.
    The semantics of these new node expressions over models $\model = \tup{N, \set{R_\pathh}_{\pathh \in \rels}, \set{\approx_{\cmp}}_{\cmp\in\cmps}, \valuation, \naming}$ is given as:
    \[
        \begin{array}{lcl}
            \model,n\vDash\exists x.\varphi & \text{\it iff } & \text{there is $ \naming'$ s.t. } \model',n\vDash \varphi,
        \end{array}
    \]
    where $\model'= \tup{N, \set{R_\pathh}_{\pathh \in \rels}, \set{\approx_{\cmp}}_{\cmp\in\cmps}, \valuation, \naming'}$ for  $\naming'$ defined exactly as $\naming$ except, perhaps, for $x$. The expression $\forall x.\varphi$ is defined as $\neg\exists.\neg\varphi$.  We will use $\exists x,y,\ldots$ and $\forall x,y,\ldots$ as shortcuts for $\exists x.\exists y\ldots$ and $\forall x.\forall y\ldots$, respectively.
\end{defi}

The strong hybrid language helps highlight a key limitation of pure axioms and motivates the need for node-creating rules. Pure axioms correspond to universally quantified expressions in the strong hybrid language (see~\cite[Prop.~3.2]{Blackburn2002BeyondPA} for details). However, certain natural frame properties cannot be captured using pure axioms alone.
A well-known example is the Church-Rosser class of frames, or the property of right-directness. The former property can be expressed as 
\[
    \forall i,j,k.\exists x. (i{:}\tup{\child} j \land  i{:}\tup{\child}x \to j{:}\tup{\child}x \land k{:}\tup{\child}x),
\]
while the latter can be expressed as
\[
    \forall i,j.\exists k. (i{:}\tup{\child} k \land j{:}\tup{\child} k).
\]
Both expressions involve a $\forall\exists$ quantification pattern, which goes beyond the purely universal pattern allowed by pure axioms. Node-creating rules provide a natural way to handle such $\forall\exists$ properties: the premises correspond to the universally quantified parts, while the conclusions create new nodes serving as witnesses for the existential quantifiers.

For example,
\[
\forall i,j.\exists k. \tup{@_i\neq_\cmp @_k}\land\tup{@_j\neq_\cmp @_k}
\]
can be casted into a node generating rule as
\[
\begin{array}{c}
{\infer[\text{with $k$ a new nominal.}]
{
\deduce[]
{\phantom{|} {\tup{@_i\neq_\cmp @_k}}}    
{\phantom{|}{\tup{@_j\neq_\cmp @_k}}}}
{\text{$i,j$ in the branch}
}
}
\end{array}
\]
Incorporating this rule into our tableaux ensures that for every pair of nodes $i$ and $j$ in a branch, there is a node $k$ whose data value differs from both.

Let us now move to the general form of these rules and how to represent them. Let $\forall x_1,\ldots,x_\ell.\exists y_1,\ldots y_{\ell'}.\varphi$ be a node expression from the strong hybrid language, such that $\varphi$ is quantifier free. For every such a node expression we define the rule  
\[
\begin{array}{c}
{\infer[\text{with $k_1,\ldots,k_{\ell'}$ new nominals.}]
{
\phantom{|} {\varphi[x_1 \leftarrow i_1,\ldots,x_\ell \leftarrow i_\ell, y_1\leftarrow k_1\ldots y_{\ell'}\leftarrow k_{\ell'}]}}
{\text{$i_1,\ldots,i_{\ell}$ in the branch}}
}
\end{array}
\]

Once more, we can prove a Strong Completeness Theorem for Node Creating Rules.

\begin{thm}[Strong Completeness for Node Creating Rules]
    Let $S$ be a set of node expressions from the strong hybrid language of the form $\forall x_1,\ldots,x_\ell.\exists y_1,\ldots y_{\ell'}.\varphi$, where $\varphi$ is quantifier free.
    Extend the tableau calculus from~\Cref{sec:tableaux} with the associated rules as defined above. The obtained tableaux is complete with respect to the class of models whose frames are defined by $\bigwedge_{\psi\in S} \psi$.
\end{thm}

\begin{proof}
The argument is as in the proof of~\Cref{th:strong}.
\end{proof}
\section{Final Remarks}
\label{sec:final}

We introduced tableau calculi for data-aware logics, with particular attention to variants of the query language XPath extended with data comparisons. Precisely, we presented an internalized tableau calculus for the logic called $\Hxpath$. This logic extends the standard XPath formalism as studied in the context of Modal Logic by incorporating data comparisons, along with nominals and satisfiability modalities from Hybrid Logic.
We proved that our tableau calculus is sound, complete, and terminating, making it a practical tool for checking the satisfiability of $\Hxpath$ node and path expressions. Additionally, we showed that rule applications can be systematically restricted to ensure that tableau construction runs in polynomial space, without sacrificing completeness. This yields a \pspace-completeness result for the satisfiability problem in $\Hxpath$.

Our work builds upon~\cite{ArecesFS17} and offers several enhancements. Most notably, it provides a detailed account of the termination argument, which was only briefly outlined in the earlier article. The presence of nominals and satisfiability modalities in the language implies that the tableau calculi need to perform equality reasoning, which complicates termination. In particular, the introduction of new nominals needs to be carefully handled.  In addition, we demonstrate that the \textsc{PSpace} upper bound also applies in the case of trees.
The current proposal is, arguably, simpler and more elegant than the presentation of the calculus given in~\cite{ArecesFS17}.  In particular, the \pspace upper bound for satisfiability is presented in full, at a level of detail suitable for implementation. The tableau calculi introduced in this article are \emph{internalized}, i.e., rules only contain expressions of $\Hxpath$. It is known that internalized proof systems for Hybrid Logics enjoy some nice properties~\cite{Blackburn00,Blackburn2002BeyondPA}. We exploited these properties here, demonstrating how to extend the tableau calculus with pure axioms and node creating rules to characterize interesting model classes. Moreover, we showed that strong completeness theorems holds for these particular kind of tableau rules, ensuring that the resulting calculi are automatically complete for the corresponding model classes.  We illustrated how these rules can be used to capture new operators and model classes relevant for applications based on semi-structured data. 


The results presented in this article pave the way to a modal investigation of other data-aware logics that are receiving considerable attention recently, such as GQL~\cite{FrancisGGLMMMPR23,FrancisGGLMMMPR23b} and SHACL~\cite{Ortiz23,Ahmetaj0S23}. It also makes it possible to consider Modal Logics built over more complex data structures, such as Concrete Domains~\cite{DemriQ21}. It would be interesting to explore futher connections with Register Automata~\cite{KaminskiF94} and similar structures. Finally, implementing a tableaux-based procedure for data-aware languages over the hybrid prover \textsf{HTab}~\cite{Hoffmann2007} is also part of our agenda.

\paragraph{\bf Acknowledgments}
This work is supported by projects ANPCyT-PICT-2020-3780, 
PICT-2021-00400, 
CONICET projects PIBAA-28720210100428CO, PIBAA28720210100165CO, 
Stic-AmSud 23-STIC-07 `DL(R)', Secyt-UNC, 
the EU Grant Agreement 101008233 (MISSION) 
and
by the Laboratoire International Associ\'e SINFIN.

\bibliography{references}

\newcommand{\etalchar}[1]{$^{#1}$}
\begin{thebibliography}{FGG{\etalchar{+}}23b}

\bibitem[ABFF16]{KR16}
S.~Abriola, P.~Barcel{\'{o}}, D.~Figueira, and S.~Figueira.
\newblock Bisimulations on data graphs.
\newblock In {\em Principles of Knowledge Representation and Reasoning: Proceedings of the Fifteenth International Conference, {KR}}, pages 309--318, 2016.

\bibitem[ABM00]{Areces2000c}
C.~Areces, P.~Blackburn, and M.~Marx.
\newblock The computational complexity of hybrid temporal logics.
\newblock {\em Logic Journal of the IGPL}, 8(5):653--679, 2000.

\bibitem[ABS99]{Abit:data99}
S.~Abiteboul, P.~Buneman, and D.~Suciu.
\newblock {\em Data on the Web: From Relations to Semistructured Data and {XML}}.
\newblock Morgan Kaufmann, 1999.

\bibitem[ACDF23]{arec:data23}
C.~Areces, V.~Cassano, D.~Dutto, and R.~Fervari.
\newblock Data graphs with incomplete information (and a way to complete them).
\newblock In {\em Logics in Artificial Intelligence. 18th European Conference, JELIA 2023}, Lecture Notes in Computer Science, pages 729--744. Springer, 2023.

\bibitem[ADF14]{ADF14}
S.~Abriola, M.~Descotte, and S.~Figueira.
\newblock Definability for downward and vertical {{XP}ath} on data trees.
\newblock In {\em 21th Workshop on Logic, Language, Information and Computation}, volume 6642 of {\em LNCS}, pages 20--34, 2014.

\bibitem[ADF17]{ADF14journal}
S.~Abriola, M.~Descotte, and S.~Figueira.
\newblock Model theory of {XPath} on data trees. {P}art {II}: Binary bisimulation and definability.
\newblock {\em Information and Computation}, 255:195--223, 2017.
\newblock URL: \url{https://www.sciencedirect.com/science/article/pii/S0890540117300032}, \href {https://doi.org/10.1016/j.ic.2017.01.002} {\path{doi:10.1016/j.ic.2017.01.002}}.

\bibitem[ADFF17]{ADFF16report}
S.~Abriola, M.~Descotte, R.~Fervari, and S.~Figueira.
\newblock Axiomatizations for downward {XP}ath on data trees.
\newblock {\em Journal of Computer and System Sciences}, 89:209--245, 2017.

\bibitem[AF16]{ArecF:hilb16}
C.~Areces and R.~Fervari.
\newblock Hilbert-style axiomatization for hybrid {XP}ath with data.
\newblock In Loizos M. and A.~Kakas, editors, {\em Logics in Artificial Intelligence - 15th European Conference, {JELIA} 2016}, volume 10021 of {\em LNCS}, pages 34--48, 2016.

\bibitem[AF21]{ArecesF21}
C.~Areces and R.~Fervari.
\newblock Axiomatizing hybrid {XP}ath with data.
\newblock {\em Logical Methods in Computer Science}, 17(3), 2021.
\newblock URL: \url{https://doi.org/10.46298/lmcs-17(3:5)2021}, \href {https://doi.org/10.46298/LMCS-17(3:5)2021} {\path{doi:10.46298/LMCS-17(3:5)2021}}.

\bibitem[AFS17]{ArecesFS17}
C.~Areces, R.~Fervari, and N.~Seiler.
\newblock Tableaux for hybrid {XP}ath with data.
\newblock In {\em Progress in Artificial Intelligence - 18th {EPIA} Conference on Artificial Intelligence, {EPIA} 2017}, volume 10423 of {\em LNCS}, pages 611--623. Springer, 2017.
\newblock \href {https://doi.org/10.1007/978-3-319-65340-2\_50} {\path{doi:10.1007/978-3-319-65340-2\_50}}.

\bibitem[AOOS23]{Ahmetaj0S23}
S.~Ahmetaj, M.~Ortiz, A.~Oudshoorn, and M.~Simkus.
\newblock Reconciling {SHACL} and ontologies: Semantics and validation via rewriting.
\newblock In {\em {ECAI} 2023 - 26th European Conference on Artificial Intelligence}, volume 372 of {\em Frontiers in Artificial Intelligence and Applications}, pages 27--35. {IOS} Press, 2023.
\newblock \href {https://doi.org/10.3233/FAIA230250} {\path{doi:10.3233/FAIA230250}}.

\bibitem[AtC06]{arec:hybr05b}
C.~Areces and B.~ten Cate.
\newblock Hybrid logics.
\newblock In P.~Blackburn, F.~Wolter, and J.~van Benthem, editors, {\em Handbook of Modal Logic}, pages 821--868. Elsevier, 2006.

\bibitem[BB07]{BolanderB07}
T.~Bolander and P.~Blackburn.
\newblock Termination for hybrid tableaus.
\newblock {\em Journal of Logic and Computation}, 17(3):517--554, 2007.

\bibitem[BdRV01]{blackburn2001modal}
P.~Blackburn, M.~de~Rijke, and Y.~Venema.
\newblock {\em Modal Logic}, volume~53 of {\em Cambridge Tracts in Theoretical Computer Science}.
\newblock Cambridge University Press, 2001.

\bibitem[BFG08]{BFG08}
M.~Benedikt, W.~Fan, and F.~Geerts.
\newblock {XP}ath satisfiability in the presence of {DTD}s.
\newblock {\em Journal of the ACM}, 55(2):1--79, 2008.
\newblock \href {https://doi.org/10.1145/1346330.1346333} {\path{doi:10.1145/1346330.1346333}}.

\bibitem[BK09]{BK08}
M.~Benedikt and C.~Koch.
\newblock {XP}ath leashed.
\newblock {\em ACM Computing Surveys}, 41(1):3:1--3:54, January 2009.
\newblock URL: \url{http://doi.acm.org/10.1145/1456650.1456653}, \href {https://doi.org/10.1145/1456650.1456653} {\path{doi:10.1145/1456650.1456653}}.

\bibitem[Bla00]{Blackburn00}
P.~Blackburn.
\newblock Internalizing labelled deduction.
\newblock {\em Journal of Logic and Computation}, 10(1):137--168, 2000.
\newblock URL: \url{https://doi.org/10.1093/logcom/10.1.137}, \href {https://doi.org/10.1093/LOGCOM/10.1.137} {\path{doi:10.1093/LOGCOM/10.1.137}}.

\bibitem[BLS16]{BaeldeLS15}
D.~Baelde, S.~Lunel, and S.~Schmitz.
\newblock A sequent calculus for a modal logic on finite data trees.
\newblock In {\em 25th {EACSL} Annual Conference on Computer Science Logic, {CSL} 2016}, pages 32:1--32:16, 2016.

\bibitem[BMSS09]{BojanczykMSS09}
M.~Boja{\'{n}}czyk, A.~Muscholl, T.~Schwentick, and L.~Segoufin.
\newblock Two-variable logic on data trees and {XML} reasoning.
\newblock {\em Journal of the ACM}, 56(3), 2009.
\newblock URL: \url{http://doi.acm.org/10.1145/1516512.1516515}, \href {https://doi.org/10.1145/1516512.1516515} {\path{doi:10.1145/1516512.1516515}}.

\bibitem[Bra07]{Brauner07}
T.~Bra{\"{u}}ner.
\newblock Why does the proof-theory of hybrid logic work so well?
\newblock {\em Journal of Applied Logic}, 17(4):521--543, 2007.
\newblock URL: \url{https://doi.org/10.3166/jancl.17.521-543}, \href {https://doi.org/10.3166/JANCL.17.521-543} {\path{doi:10.3166/JANCL.17.521-543}}.

\bibitem[Bra11]{Brauner2011}
T.~Bra\"uner.
\newblock {\em Hybrid Logic and its Proof-Theory}, volume~37 of {\em Applied Logics Series}.
\newblock Springer, 2011.

\bibitem[BS95]{BlackburnS95}
P.~Blackburn and J.~Seligman.
\newblock Hybrid languages.
\newblock {\em Journal of Logic, Language and Information}, 4(3):251--272, 1995.
\newblock \href {https://doi.org/10.1007/BF01049415} {\path{doi:10.1007/BF01049415}}.

\bibitem[BtC02]{Blackburn2002BeyondPA}
P.~Blackburn and B.~ten Cate.
\newblock Beyond pure axioms: Node creating rules in hybrid tableaux.
\newblock In {\em 4th Workshop on Hybrid Logics (Hylo@LICS)}, 2002.

\bibitem[BtC06]{BtC06}
P.~Blackburn and B.~ten Cate.
\newblock Pure extensions, proof rules, and hybrid axiomatics.
\newblock {\em Studia Logica}, 84(2):277--322, 2006.

\bibitem[Bun97]{Bune:semi97}
P.~Buneman.
\newblock Semistructured data.
\newblock In {\em In {ACM} {S}ymposium on Principles of Database Systems ({PODS}'97)}, pages 117--121, 1997.

\bibitem[BvB06]{blackburn06}
P.~Blackburn and J.~van Benthem.
\newblock Modal {L}ogic: {A} {S}emantic {P}erspective.
\newblock In P.~Blackburn, F.~Wolter, and J.~van Benthem, editors, {\em Handbook of Modal Logic}, pages 1--84. Elsevier, 2006.

\bibitem[BWvB06]{handbook06}
P.~Blackburn, F.~Wolter, and J.~van Benthem, editors.
\newblock {\em Handbook of Modal Logic}.
\newblock Elsevier, 2006.

\bibitem[CD99]{xpath:w3c}
J.~Clark and S.~DeRose.
\newblock {XML} path language ({{XP}ath}).
\newblock Website, 1999.
\newblock {W3C Recommendation}. \url{http://www.w3.org/TR/xpath}.

\bibitem[Cla99]{xslt:w3c}
J~Clark.
\newblock {XSL} transformations ({XSLT}).
\newblock Website, 1999.
\newblock {W3C Recommendation}. \url{http://www.w3.org/TR/xslt}.

\bibitem[CLM10]{cateLM10}
B.~ten Cate, T.~Litak, and M.~Marx.
\newblock Complete axiomatizations for {{XP}ath} fragments.
\newblock {\em Journal of Applied Logic}, 8(2):153--172, 2010.
\newblock URL: \url{http://dx.doi.org/10.1016/j.jal.2009.09.002}, \href {https://doi.org/10.1016/j.jal.2009.09.002} {\path{doi:10.1016/j.jal.2009.09.002}}.

\bibitem[CM09]{CateM09}
B.~ten Cate and M.~Marx.
\newblock Axiomatizing the logical core of {{XP}ath} 2.0.
\newblock {\em Theory of Computing Systems}, 44(4):561--589, 2009.
\newblock URL: \url{http://dx.doi.org/10.1007/s00224-008-9151-9}, \href {https://doi.org/10.1007/s00224-008-9151-9} {\path{doi:10.1007/s00224-008-9151-9}}.

\bibitem[DQ21]{DemriQ21}
S.~Demri and K.~Quaas.
\newblock Concrete domains in logics: a survey.
\newblock {\em {ACM} {SIGLOG} News}, 8(3):6--29, 2021.
\newblock \href {https://doi.org/10.1145/3477986.3477988} {\path{doi:10.1145/3477986.3477988}}.

\bibitem[FFA14]{ICDT14}
D.~Figueira, S.~Figueira, and C.~Areces.
\newblock Basic model theory of {{XP}ath} on data trees.
\newblock In {\em International Conference on Database Theory}, pages 50--60, 2014.

\bibitem[FFA15]{ICDT14Jair}
D.~Figueira, S.~Figueira, and C.~Areces.
\newblock Model theory of {{XP}ath} on data trees. {Part~I}: Bisimulation and characterization.
\newblock {\em Journal of Artificial Intelligence Research}, 53:271--314, 2015.

\bibitem[FGG{\etalchar{+}}23a]{FrancisGGLMMMPR23}
N.~Francis, A.~Gheerbrant, P.~Guagliardo, L.~Libkin, V.~Marsault, W.~Martens, F.~Murlak, L.~Peterfreund, A.~Rogova, and D.~Vrgoc.
\newblock {GPC:} {A} pattern calculus for property graphs.
\newblock In {\em 42nd {ACM} {SIGMOD-SIGACT-SIGAI} Symposium on Principles of Database Systems, ({PODS}'23)}, pages 241--250. {ACM}, 2023.
\newblock \href {https://doi.org/10.1145/3584372.3588662} {\path{doi:10.1145/3584372.3588662}}.

\bibitem[FGG{\etalchar{+}}23b]{FrancisGGLMMMPR23b}
N.~Francis, A.~Gheerbrant, P.~Guagliardo, L.~Libkin, V.~Marsault, W.~Martens, F.~Murlak, L.~Peterfreund, A.~Rogova, and D.~Vrgoc.
\newblock A researcher's digest of {GQL} (invited talk).
\newblock In {\em 26th International Conference on Database Theory, {ICDT} 2023}, volume 255 of {\em LIPIcs}, pages 1:1--1:22. Schloss Dagstuhl - Leibniz-Zentrum f{\"{u}}r Informatik, 2023.
\newblock \href {https://doi.org/10.4230/LIPICS.ICDT.2023.1} {\path{doi:10.4230/LIPICS.ICDT.2023.1}}.

\bibitem[Fig10]{FigPhD}
D.~Figueira.
\newblock {\em Reasoning on Words and Trees with Data}.
\newblock Ph{D} thesis, Laboratoire Sp{\'e}cification et V{\'e}rification, ENS Cachan, France, 2010.
\newblock URL: \url{http://www.lsv.ens-cachan.fr/~figueira/phd/latest/thesisFigueira.pdf}.

\bibitem[Fig12]{Figueira12ACM}
D.~Figueira.
\newblock Decidability of downward {{XP}ath}.
\newblock {\em {ACM} Transactions on Computational Logic}, 13(4):34, 2012.
\newblock URL: \url{http://doi.acm.org/10.1145/2362355.2362362}, \href {https://doi.org/10.1145/2362355.2362362} {\path{doi:10.1145/2362355.2362362}}.

\bibitem[Fig13]{Fig13}
D.~Figueira.
\newblock On {XP}ath with transitive axes and data tests.
\newblock In Wenfei Fan, editor, {\em In {ACM} {S}ymposium on Principles of Database Systems ({PODS}'13)}, pages 249--260, New York, NY, USA, 2013. ACM Press.
\newblock \href {https://doi.org/10.1145/2463664.2463675} {\path{doi:10.1145/2463664.2463675}}.

\bibitem[Fit83]{Fit83}
M.~Fitting.
\newblock {\em Proof Methods for Modal and Intuitionistic Logics}.
\newblock Synthese Library Volume. Springer Netherlands, 1983.
\newblock URL: \url{https://books.google.com.ar/books?id=FfQOdQssjCAC}.

\bibitem[Fit96]{Fit96}
M.~Fitting.
\newblock {\em First-Order Logic and Automated Theorem Proving, Second Edition}.
\newblock Graduate Texts in Computer Science. Springer, 1996.
\newblock \href {https://doi.org/10.1007/978-1-4612-2360-3} {\path{doi:10.1007/978-1-4612-2360-3}}.

\bibitem[FS11]{FigueiraS11}
D.~Figueira and L.~Segoufin.
\newblock Bottom-up automata on data trees and vertical {XP}ath.
\newblock In {\em 28th International Symposium on Theoretical Aspects of Computer Science ({STACS} 2011)}, pages 93--104, 2011.
\newblock URL: \url{http://dx.doi.org/10.4230/LIPIcs.STACS.2011.93}, \href {https://doi.org/10.4230/LIPIcs.STACS.2011.93} {\path{doi:10.4230/LIPIcs.STACS.2011.93}}.

\bibitem[GA21]{GonzalezA21}
N.~Gonz{\'{a}}lez and S.~Abriola.
\newblock Characterizations for {XP}ath$_\mathbf{R}(\downarrow)$.
\newblock In A.~Silva, R.~Wassermann, and R.~de~Queiroz, editors, {\em Logic, Language, Information, and Computation - 27th International Workshop, WoLLIC 2021}, volume 13038 of {\em LNCS}, pages 319--336. Springer, 2021.
\newblock \href {https://doi.org/10.1007/978-3-030-88853-4\_20} {\path{doi:10.1007/978-3-030-88853-4\_20}}.

\bibitem[GKP05]{GKP05}
G.~Gottlob, C.~Koch, and R.~Pichler.
\newblock Efficient algorithms for processing {{XP}ath} queries.
\newblock {\em ACM Transactions on Database Systems}, 30(2):444--491, 2005.
\newblock \href {https://doi.org/10.1145/1071610.1071614} {\path{doi:10.1145/1071610.1071614}}.

\bibitem[HA09]{Hoffmann2007}
G.~Hoffmann and C.~Areces.
\newblock {HTab}: A terminating tableaux system for hybrid logic.
\newblock {\em Electronic Notes in Theoretical Computer Science}, 231:3--19, March 2009.
\newblock \href {https://doi.org/10.1016/j.entcs.2009.02.026} {\path{doi:10.1016/j.entcs.2009.02.026}}.

\bibitem[KF94]{KaminskiF94}
M.~Kaminski and N.~Francez.
\newblock Finite-memory automata.
\newblock {\em Theoretical Computer Science}, 134(2):329--363, 1994.
\newblock \href {https://doi.org/10.1016/0304-3975(94)90242-9} {\path{doi:10.1016/0304-3975(94)90242-9}}.

\bibitem[Lad77]{Ladner77}
R.~Ladner.
\newblock The computational complexity of provability in systems of modal propositional logic.
\newblock {\em {SIAM} Journal of Computing}, 6(3):467--480, 1977.
\newblock \href {https://doi.org/10.1137/0206033} {\path{doi:10.1137/0206033}}.

\bibitem[Mar04]{M04}
M.~Marx.
\newblock {XP}ath with conditional axis relations.
\newblock In {\em International Conference on Extending Database Technology (EDBT'04)}, volume 2992 of {\em LNCS}, pages 477--494. Springer, 2004.
\newblock \href {https://doi.org/10.1007/b95855} {\path{doi:10.1007/b95855}}.

\bibitem[Mar06]{Marx06}
M.~Marx.
\newblock Complexity of modal logic.
\newblock In P.~Blackburn, F.~Wolter, and J.~van Benthem, editors, {\em Handbook of Modal Logic}, pages 139--179. Elsevier, 2006.

\bibitem[MdR05]{MdR05}
M.~Marx and M.~de~Rijke.
\newblock Semantic characterizations of navigational {XP}ath.
\newblock {\em ACM SIGMOD Record}, 34(2):41--46, 2005.

\bibitem[Ort23]{Ortiz23}
M.~Ortiz.
\newblock A short introduction to {SHACL} for logicians.
\newblock In {\em Logic, Language, Information, and Computation - 29th International Workshop, WoLLIC 2023}, volume 13923 of {\em LNCS}, pages 19--32. Springer, 2023.
\newblock \href {https://doi.org/10.1007/978-3-031-39784-4\_2} {\path{doi:10.1007/978-3-031-39784-4\_2}}.

\bibitem[Pri08]{Priest08}
Graham Priest.
\newblock {\em An Introduction to Non-Classical Logic: From If to Is}.
\newblock Cambridge University Press, 2 edition, 2008.

\bibitem[RCDS14]{xquery:w3c}
J.~Robie, D.~Chamberlin, M.~Dyck, and J.~Snelson.
\newblock {XQuery} 3.0: An {XML} query language.
\newblock Website, 2014.
\newblock {W3C Recommendation}. \url{https://www.w3.org/TR/xquery-30/}.

\bibitem[Smu68]{Smullyan1968}
R.~Smullyan.
\newblock Analytic tableaux.
\newblock In {\em First-Order Logic}, pages 15--30. Springer Berlin Heidelberg, Berlin, Heidelberg, 1968.
\newblock \href {https://doi.org/10.1007/978-3-642-86718-7_2} {\path{doi:10.1007/978-3-642-86718-7_2}}.

\end{thebibliography}
\bibliographystyle{alphaurl}

\end{document}